\documentclass[prb,preprintnumbers,superscriptaddress,showpacs,twocolumn]{revtex4-1}
\usepackage{graphicx}
\usepackage{bm}
\usepackage{amsmath,amssymb,amsfonts,textcomp}
\usepackage{multirow}
\usepackage{color}
\usepackage{array}
\usepackage{hhline}
\usepackage[normalem]{ulem}

\usepackage{amsmath}    % need for subequations
\usepackage{verbatim}   % useful for program listings
\usepackage{subfigure}  % use for side-by-side figures
\begin{document}

%\title{The dynamics of superfluid $^4$He in a large energy, wave vector and pressure range} 
\title{Microscopic dynamics of superfluid $^4$He: a comprehensive study \\
by inelastic neutron scattering} 
\author{K. Beauvois} 
   \affiliation{Institut Laue-Langevin, CS 20156, 38042 Grenoble Cedex 9, France}
  \affiliation{Univ. Grenoble Alpes, CNRS, Grenoble INP, Institut N\'eel, 38000 Grenoble, France}
%\author{C. E. Campbell}
%  \affiliation{School of Physics and Astronomy, University of Minnesota, Minneapolis MN 55455, USA}
\author{J. Dawidowski}
   \affiliation{Comisi\'on Nacional de Energ\'ia At\'omica and CONICET, Centro At\'omico Bariloche, (8400) San Carlos de Bariloche,
R\'io Negro, Argentina}
\author{B. F\aa k} 
   \affiliation{Institut Laue-Langevin, CS 20156, 38042 Grenoble Cedex 9, France}
\author{H. Godfrin}
 \affiliation{Univ. Grenoble Alpes, CNRS, Grenoble INP, Institut N\'eel, 38000 Grenoble, France}
\author{E.~Krotscheck}
   \affiliation{Department of Physics, University at Buffalo, SUNY Buffalo NY 14260, USA}
   \affiliation{Institute for Theoretical Physics, Johannes Kepler University, A 4040 Linz, Austria}
%\author{H.-J. Lauter}
%   \affiliation{Spallation Neutron Source, Oak Ridge National Laboratory, Oak Ridge, Tennessee 37831, USA}
\author{J. Ollivier}
   \affiliation{Institut Laue-Langevin, CS 20156, 38042 Grenoble Cedex 9, France}\author{A. Sultan}
 \affiliation{Univ. Grenoble Alpes, CNRS, Grenoble INP, Institut N\'eel, 38000 Grenoble, France}

%\footnote{* Institute of Engineering Univ. Grenoble Alpes}

\date{\today}

\begin{abstract}

The dynamic structure factor of superfluid $^4$He has been 
investigated at very low temperatures by inelastic neutron scattering. 
The measurements combine different incoming energies resulting in an unprecedentedly large dynamic range with excellent energy resolution,
covering wave vectors $Q$ up to 5\,\AA$^{-1}$ and energies 
$\omega$ up to 15\,meV.
A detailed description of the dynamics of
superfluid $^4$He is obtained  
from saturated vapor pressure up to solidification. 
The single-excitation spectrum is substantially
modified at high pressures, as the maxon energy exceeds the roton-roton decay threshold. 
A highly structured multi-excitation spectrum is observed at low
energies, where clear thresholds and branches have been identified.
Strong phonon emission branches are observed when the phonon 
or roton group velocities exceed the sound velocity. 
The spectrum is found to display strong multi-excitations whenever the
single-excitations face disintegration following Pitaevskii's type a
or b criteria.  At intermediate energies, an interesting pattern in
the dynamic structure factor is observed in the vicinity of the recoil
energy.  All these features, which evolve significantly with pressure, 
are in very good agreement with the Dynamic Many-body calculations, 
even at the highest densities, where the correlations are strongest.

\end{abstract}

%\pacs{05.30.Jp,67.25.dt,67.25.D-,78.70.Nx}

\maketitle 

\section{Introduction}
Understanding the dynamics of correlated bosons is a subject of
general interest in several fields of physics. Bose-Einstein
condensation and superfluidity \cite{PinesNoz,leggett2006quantum},
first found in $^4$He, are fundamental phenomena that imprint
remarkable signatures on the dynamics of these systems.
Experimentally, superfluid $^4$He is the simplest example of strongly
correlated bosons. The interaction potential is particularly well
known, and substantial effort has been devoted to develop a coherent
theoretical framework able to describe and explain the extraordinary
properties of this quantum fluid
\cite{PinesNoz,leggett2006quantum,Landauroton2,GlydeBook,TriesteBook,eomI,Vitali,roggero2013dynamical,eomIII,ferre2016dynamic,dmowski2017}. 
The theoretical methods can be generalized to other many-body problems,
including for instance up-to-date approaches of the complex case of
correlated fermions
\cite{polish,PRB2010dynamic,Nature_2p2h,nava2013dynamic,gordillo2016liquid}.
% include EK, JLTP 119, "103-145" (2000)?

%\section{The dynamic structure factor}
The prediction by Landau \cite{Landauroton2} of the phonon-roton
excitation spectrum of superfluid $^4$He and its direct observation in
the dynamic structure factor $S(Q,\omega)$ using neutron scattering
techniques \cite{GlydeBook,Glyde2018review} are cornerstones of modern
physics, at the origin of the present microscopic 
descriptions of matter \cite{anderson1997concepts,TriesteBook,fetter2012quantum}.
The dynamics of superfluid $^4$He at very low temperatures, in the
vicinity of the ground state, is dominated by the 
``phonon-maxon-roton'' excitation branch. 
The corresponding excitations, extremely sharp, correspond essentially to poles of the dynamic density-density response function. They 
are referred to as ``single-excitations'' in the neutron scattering literature, 
and as ``quasi-particles'' in theoretical works. 
An effective description of the dynamics of such systems can be obtained 
in terms of these modes, allowing for instance a very accurate 
statistical evaluation of the low temperature thermodynamic properties
\cite{GlydeBook,DonnellyDonnellyHills}.

Sharp excitations are absent above twice the roton energy
\cite{GlydeBook,Glyde2018review,pitaevskii59}, and the dynamics at
intermediate energies is described in terms of broad excitations, 
named ``multi-excitations'' for reasons described below.
Multi-excitations still have a significant statistical weight in the dynamic
structure factor
\cite{GlydeBook,Glyde2018review,CowleyWoods,Svensson1976,Andersen92,Andersen94a,Crevecoeur,GibbsThesis,Gibbs1999,Gibbs2000,Ketty}.
Their spectrum is known to display some structure since
the early measurements of Svensson, Martel, Sears and Woods
\cite{Svensson1976}.  More recent investigations 
\cite{Andersen92,Andersen94a,Crevecoeur,GibbsThesis,Gibbs1999,Gibbs2000}
showed that some features could be ascribed to
multi-excitations. These were related to pairs of high
density-of-states roton (R) and maxon (M) modes (noted hereafter 2R,
2M, and MR). The broad ridges observed in $S(Q,\omega)$ at SVP (see
Figure\,1 of Ref.\,\onlinecite{Andersen92}), and at 20 bars (see
Figure\,1 of Ref.\,\onlinecite{Gibbs2000}) were consistent with the
calculated energies of the main combinations (2R, 2M and MR).

A much finer structure in the dynamic response was observed in our 
recent work at zero pressure \cite{Ketty}, including sharp thresholds, 
narrow branches, and a new two-phonon decay process, the ``ghost phonon''.
Explaining this rich dynamic response, observed from the continuum 
limit to subatomic distances, constitutes a challenge and an opportunity 
for microscopic theories. 

Finally, at high energies, the dynamic structure factor gradually
approaches a quasi-free-particle behavior \cite{CowleyWoods} described
by the impulse approximation
\cite{GlydeBook,Glyde2018review,Prisk2017}.

\begin{figure*}[t]
	\includegraphics[width=0.95\textwidth]{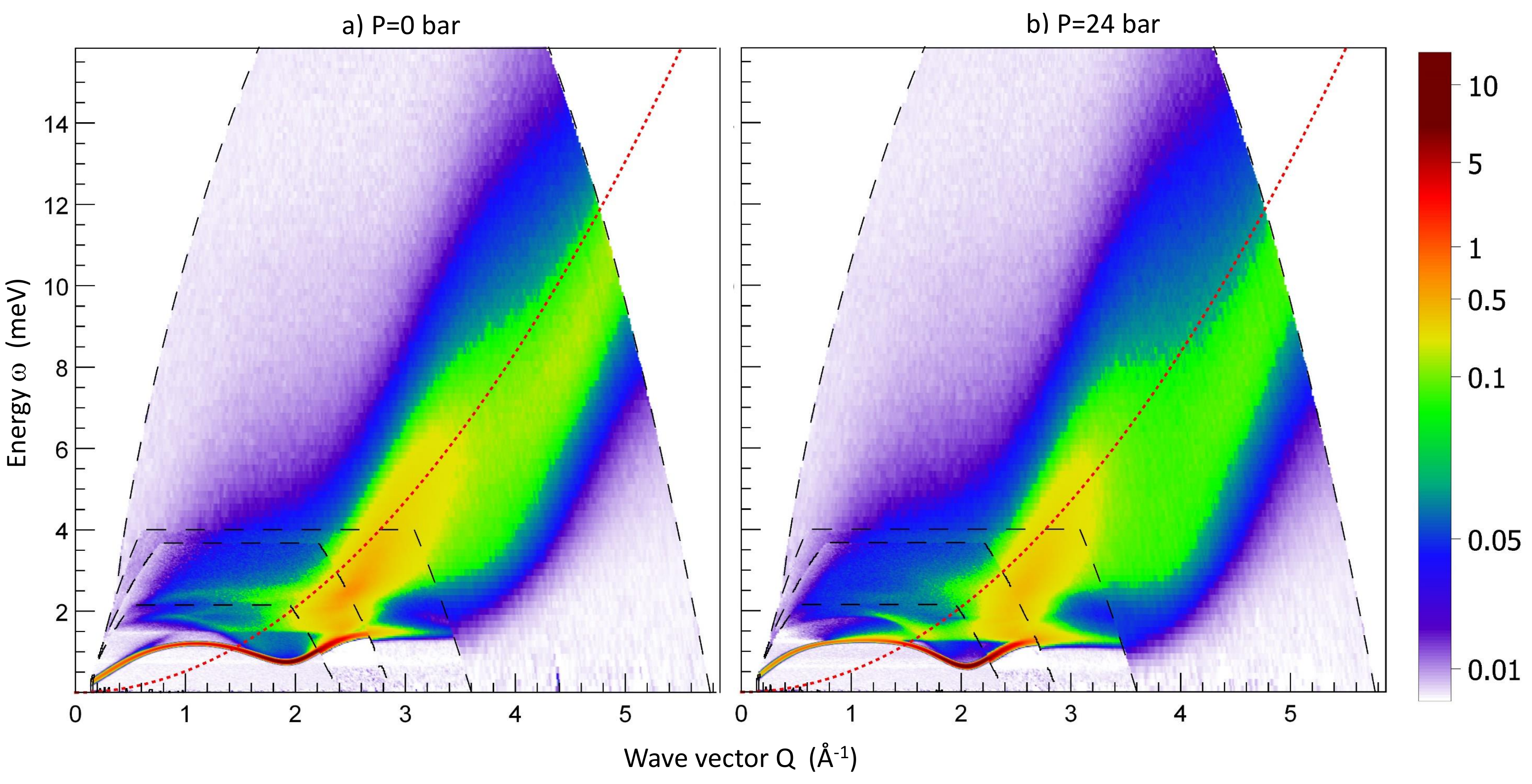}
\caption{$S(Q,\omega)$ of superfluid $^4$He as a function of wave
  vector and energy transfer, measured at $T\le100$\,mK at (a)
  saturated vapor pressure ($P\approx0$) and (b) near solidification
  ($P=24$\,bars).  The plots combine data measured at different incident
  neutron energies ($E_i$=3.55, 5.1, 8.00 and 20.45\,meV) for an
  optimum energy resolution; the dashed black lines represent the
  limits of the corresponding kinetic ranges.  The dotted red line is
  the free $^4$He atom recoil energy $E_r=\frac{\hbar^2 Q^2}{2M}$.  The
  color-coded intensity scale is in units of meV$^{-1}$.  }
\label{completemap}
\end{figure*}

Even though helium is one of the most intensively investigated
physical substances, measurements covering a large kinetic range are
scarce. The canonical results by Cowley and Woods \cite{CowleyWoods},
Dietrich {\em et al.}\cite{dietrich72} or Svensson {\em et
al.}\cite{Svensson1976} have a low resolution by modern standards,
while later measurements specialize in specific ranges
\cite{Graf74,Stirling91,GibbsThesis,Gibbs1999,Gibbs2000}. Our extensive
high-resolution measurements, presented in Fig.\,\ref{completemap},
provide a detailed and complete map of the dynamics of superfluid
$^4$He. In addition to its aesthetic merits, the picture shows new
features which are the object of this manuscript.

Helium is highly compressible. Since the atomic correlations depend on
the density, it is interesting to investigate the pressure dependence
of the density excitations. Much of the earlier work has been focused
on the effect of pressure on the single-excitation response, in order
to determine, for example, the Landau parameters characterizing the
dispersion curve.  The multi-excitation spectrum has also been found
experimentally
\cite{dietrich72,Graf74,Stirling91,GibbsThesis,Gibbs1999,Gibbs2000,Pearce2001} 
and theoretically \cite{GlydeBook,Manousakis84,Vitali,eomIII} to be
strongly modified by the pressure. It was therefore desirable to
extend our recent high sensitivity measurements \cite{Ketty} to finite
pressures, and more particularly near solidification, where theory
\cite{eomIII} predicted radical changes in the dynamics.

In this manuscript, we present a detailed investigation of the effect
of pressure on the dynamics of superfluid $^4$He.  We cover a large
energy and wave vector range while preserving the resolution needed to
observe the fine structure of the spectra.  High resolution maps of
the dynamic structure factor $S(Q,\omega)$ have been obtained at
saturated vapor pressure (SVP) and at $P=24$\,bars, close to 
solidification, as shown in Fig.\,\ref{completemap}. Additional
measurements have been made in a smaller dynamic range at the
intermediate pressures 5 and 10\,bars.  We finally compare our data to
microscopic calculations of $S(Q,\omega)$ within the Dynamic Many-Body
theory \cite{eomIII} performed at the densities corresponding to the
experimental pressure conditions.

\section{Experimental details}
\label{sec:expdetails}
The measurements were performed on the IN5 time-of-flight 
spectrometer at the high-flux reactor of Institut Laue Langevin.  
Our previous work \cite{Ketty} at low temperatures and 
saturated vapor pressure
used cold neutrons of energy $E_i$=3.55\,meV.  
In the present work, we
combine data taken using four different incident neutron energies,
$E_i$=3.55, 5.11, 8.00, and 20.45\,meV,
for which the energy resolution (FWHM) at elastic energy transfer 
was 0.070, 0.12, 0.23 and 0.92\,meV, respectively.  
This allowed us to obtain a complete map of the dynamic 
structure factor at the most
relevant pressures, {\em i.e.\/}, saturated vapor pressure (SVP) and near
solidification ($P=24$\,bars).
We also investigated a few intermediate pressures
using $E_i$=3.55\,meV.

The cylindrical sample cell was made out of aluminum 5083, with 1\,mm
wall thickness and 15\,mm inner diameter \cite{Ketty}.  Cadmium disks
of 0.5\,mm thickness were placed inside the cell every centimeter to
reduce multiple scattering.  The cell was thermally connected to the
mixing chamber of a very low temperature dilution refrigerator using
massive OFHC copper pieces.  Heat exchangers made out of sintered
silver powder were used to provide a good thermal contact with the
helium sample.  Care was taken to thermally anchor the filling
capillary at several places along the dilution unit, in order to
reduce heat leaks to the cell.  The measurements were all performed at
very low temperatures, well below 100\,mK, {\em i.e.\/} essentially at zero
temperature for the properties under investigation.

High purity (99.999\,\%) helium gas was condensed in the cell at low
temperatures, using a gas handling system including a ``dipstick'' cold
trap operated in a helium storage dewar.  The dipstick was used to
condense the gas and to pressurize the helium sample.  The pressure in
the system was measured with a precision of 6\,mbars with a 0-60\,bars
Digiquartz gauge located at the top of the cryostat.  The
corresponding precision for the pressures inside the cell is
20\,mbars, after applying helium hydrostatic head corrections.  The
actual pressures in the cell for the nominal 0, 5, 10 and 24\,bars are
essentially 0 (SVP at 100\,mK), 5.01(2), 10.01(2) and 24.08(2)\,bars.

\section{Data reduction}
\label{sec:data-reduction}

Standard time-of-flight data-reduction \cite{lamp} was 
used to obtain the dynamic structure factor $S(Q,\omega)$ from the raw
data.  The contribution of the cell scattering was subtracted, as
well as that of double scattering events of type ``inelastic helium
scattering plus elastic scattering from the cell''. This type of double
scattering is essentially independent of wave vector.

The contribution of the multiple scattering within the helium was
corrected using Monte Carlo simulations \cite{Javier}.  Due to the
small diameter of our sample cell and the presence of several cadmium
plates, multiple scattering corrections are small (the ratio of
double-scattered to single-scattered neutrons is on the order one
percent \cite{Ketty}), but may be comparable to the multi-excitation
signal. It is therefore essential to verify that multiple scattering
is not contaminating the spectra in the energy and wave vector regions
of interest, and perform the corrections when necessary, in particular
at low $Q$.

\begin{figure}[b]
	\includegraphics[width=1\columnwidth]{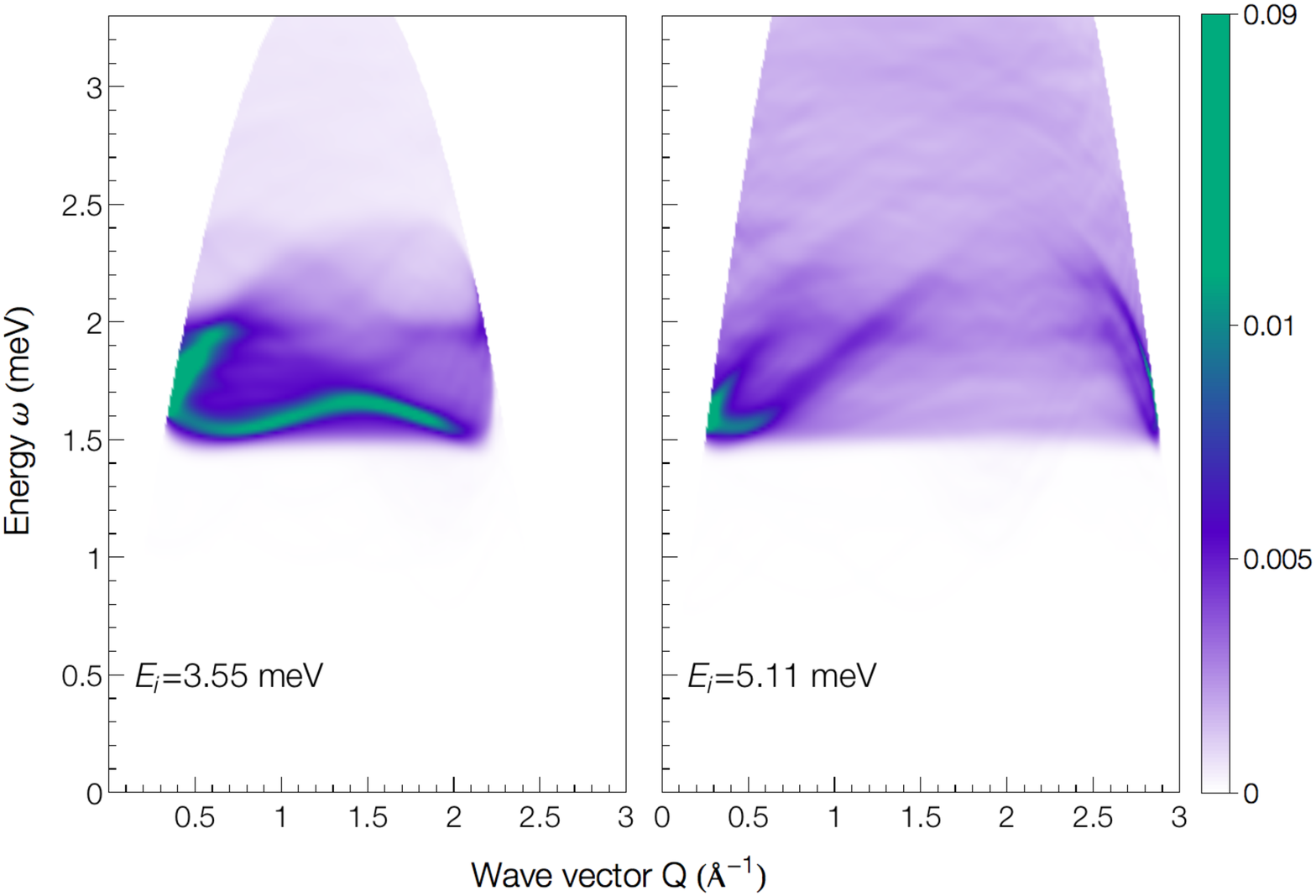}
\caption{Monte Carlo calculation of the contribution of
  double-scattering within the helium to $S(Q,\omega)$.  Results are
  shown for two incident neutron energies, $E_i$=3.55 and 5.11\,meV.  The color-coded intensity
  scale is in units of meV$^{-1}$.}
	\label{montecarlo}
\end{figure}

Since multiple scattering depends on the incident neutron energy, as shown
in figure \ref{montecarlo}, while multi-excitations do not, Monte Carlo calculations can
be used to select the most appropriate incident neutron energy for the
experiments, and also to experimentally distinguish multi-excitations
from multiple scattering.

The only input needed by the Monte Carlo simulations \cite{Javier} 
in the present case is
the initially measured scattering function $S(Q,\omega)$ after
corrections for multiple scattering processes involving the cell, and
the coherent scattering cross section of $^4$He, $\sigma_{c}=1.34$\,barns.  We
first calculate the total scattering cross section \cite{Javier,Daw1999} $\sigma_s(E_i)$:

\begin{equation}
\sigma_s(E_i)=\frac{N\sigma_{c}}{2k_i^2}\int QdQ\int S(Q,\omega)d\omega,
\end{equation}
where $N$ is the number of scatterers and $k_i$ the incident neutron
wave vector.

We find $\sigma_s(E_i=3.55\,$meV$)$=0.64 barns, about one half of the
coherent scattering cross section $\sigma_{\rm coh}$.  The multiple
scattering fraction is 0.8\,\% for $E_i$=3.55\,meV, increasing
slightly with pressure from 0.79\,\% at SVP to 1.06\,\% at
24\,bars. This agrees well with calculations using the semi-analytical
method developed by Sears \cite{Sears75}, which give values increasing
from 0.93\,\% to 1.09\,\% for the same pressures.  Multiple scattering
can be seen in the experimental spectra at low wave vectors, thus
providing a way to check the Monte Carlo calculations used to
eliminate this effect. This is a crucial step in the data analysis,
needed to ensure that all the features we report in $S(Q,\omega)$ do
indeed correspond to multi-excitations.

The calculated contribution due to multiple scattering within the
helium has been subtracted from the spectra measured using incident
neutron energies $E_i$=3.55 and 5.11\,meV. This was found to be
unnecessary for $E_i$=8.00 and 20.45\,meV, because multiple scattering
processes are negligible in the corresponding regions of
the``combined" spectra of Fig.\,\ref{completemap}.

An overall scale factor was applied to $S(Q,\omega)$ at SVP, so that
the weight of the single excitation $Z(Q)$ agrees with that of
Cowley and Woods \cite{CowleyWoods} near the roton, {\em i.e.\/},
$Z(Q=2\,$\AA$^{-1}$)=0.93 at SVP. At higher pressures, the same scaling
factor was used, but corrected for the density ratio
$\rho(P)/\rho(P=0)$.

\section{Experimental results}

\subsection{Spectra at SVP and P=24 bars in a large dynamic range}

Our comprehensive results on the dynamic structure factor
$S(Q,\omega)$ at SVP and $P=24$\,bars are shown in
Fig.\,\ref{completemap}. 
These maps were obtained by combining the four 
different neutron energies. Higher
energies make a larger dynamic range accessible, but the instrumental
energy resolution deteriorates rapidly (see section
\ref{sec:expdetails}). Since the corresponding dynamic ranges have a
substantial overlap, we can select the most appropriate data set in
terms of resolution, neutron counts or cleanest background for each
region of the $Q-\omega$ plane. The $S(Q,\omega)$ maps are built in
the following way: first, the spectrum measured at $E_i$=3.55\,meV is
represented; outside its useful kinetic range, the data at
$E_i$=5.11\,meV are added, then the data at $E_i$=8.00\,meV and
finally, the data at $E_i$= 20.45\,meV.

\begin{figure}[t]
	\includegraphics[width=1.0\columnwidth]{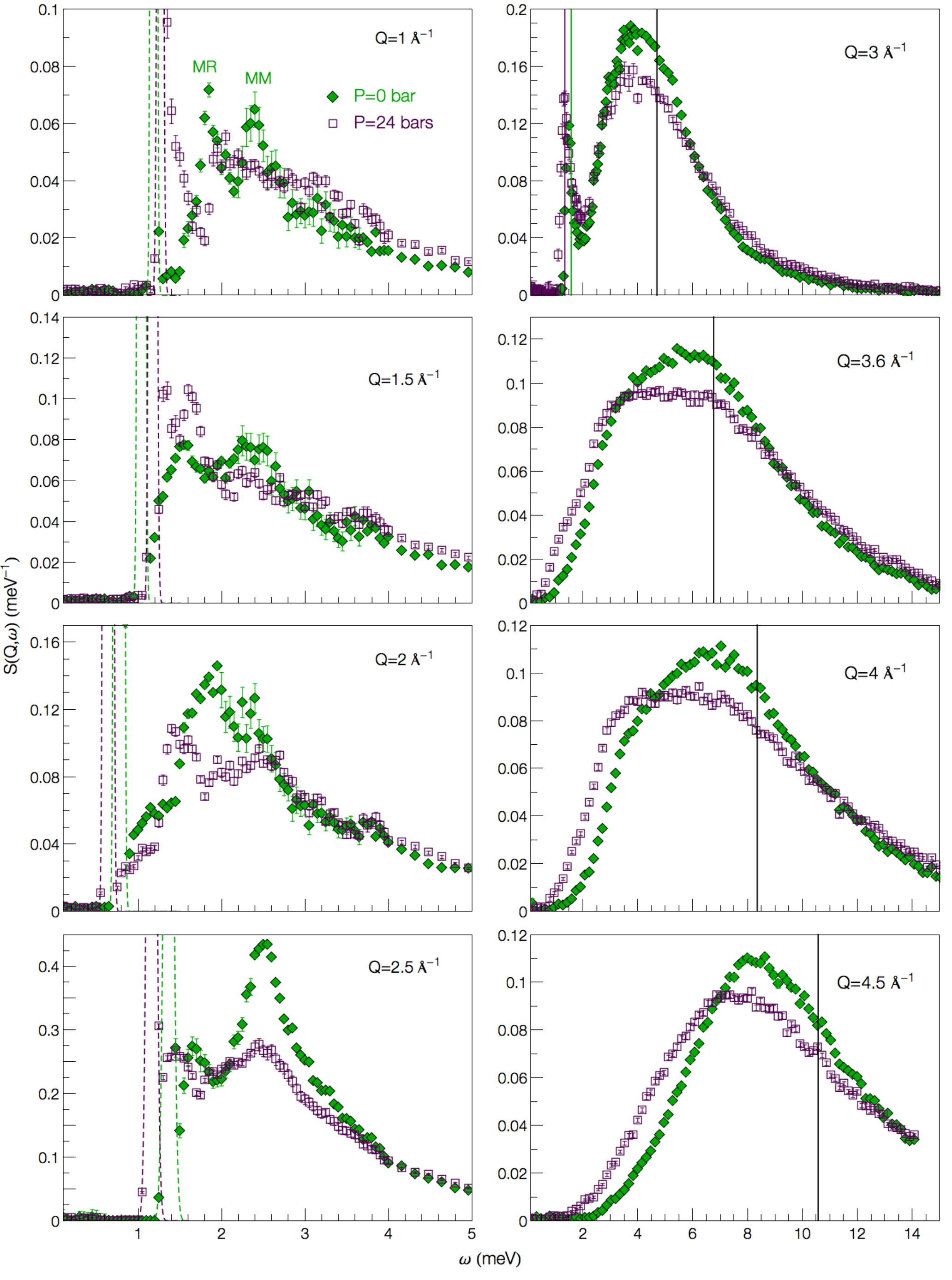}
\caption{Dynamic structure factor $S(Q,\omega)$ combining data at four
  incident neutron energies: spectra for different wave vectors $Q$ at
  SVP (green diamonds) and $P=24$ bars (purple squares). The dashed lines are
  Gaussian fits of the resolution-limited phonon-roton peaks (off
  scale).  The black lines represent the helium recoil energy.  At
  $Q=3$\,\AA$^{-1}$, the purple and green lines represent the two-roton
  energy 2$\Delta_R$ at SVP and $P=24$ bars, respectively. At
  $Q=1$\,\AA$^{-1}$, MR and MM are the energy positions at SVP of the
  maxon-roton and maxon-maxon multi-excitations, respectively.}
		\label{Cuts}
\end{figure}

The constant wave vector scans presented in Fig.\,\ref{Cuts}, obtained
as particular ``cuts'' of Fig.\,\ref{completemap}, provide a
complementary perspective on the data.  The phonon-roton
single-excitation mode is very narrow at the scale of
Figs.\,\ref{completemap} and \ref{Cuts}, and the observed width is
essentially a measure of the experimental energy resolution (with the
remarkable exception of the maxon at high pressures, which is
discussed in the next section). The influence of a finite energy
resolution is clearly seen in Fig.\,\ref{completemap} as a width
discontinuity in the Pitaevskii plateau \cite{GlydeBook,pitaevskii59},
between ranges corresponding to different incident neutron
energies. It is important to note, however, that the experimental
broadening effects are negligible in all the {\it{multi-excitation\/}}
region investigated in the present work (except at the end of  
the Pitaevskii plateau).

Merging data measured with different resolutions has been successfully
achieved, judging from the remarkable continuity in intensity between
the different regions represented in Fig.\,\ref{completemap}. This is
essentially due to the fact that the sharpest multi-excitations are
found in the low energy and low wave vector sector, adequately covered
by our high resolution data at $E_i$=3.55 and 5.11\,meV. Conversely,
the spectra in the quasi-free particle region, at high energies and
wave-vectors, are intrinsically broad, and adequately covered by our
data at 8.00 and 20.45\,meV, in spite of their lower resolution. Using
optimized incident neutron energies reveals the complete evolution of
the system, characterized by several multi-excitation branches merging
progressively at high wave vectors to form a broad but rather intense
feature.  Intensity in this region was observed in early studies
\cite{GlydeBook,Glyde2018review}, but the data where either strongly
truncated \cite{Andersen92,Andersen94a,Gibbs2000}, or measured with 
low resolution \cite{CowleyWoods}.  This feature finally
becomes, after a strong oscillation, a less intense branch
progressively approaching the free particle parabolic dispersion.

\subsection{High resolution spectra as a function of pressure}
\label{subsec:effect-of-pressure}
\begin{figure}[t!]
	\includegraphics[width=1.0\columnwidth]{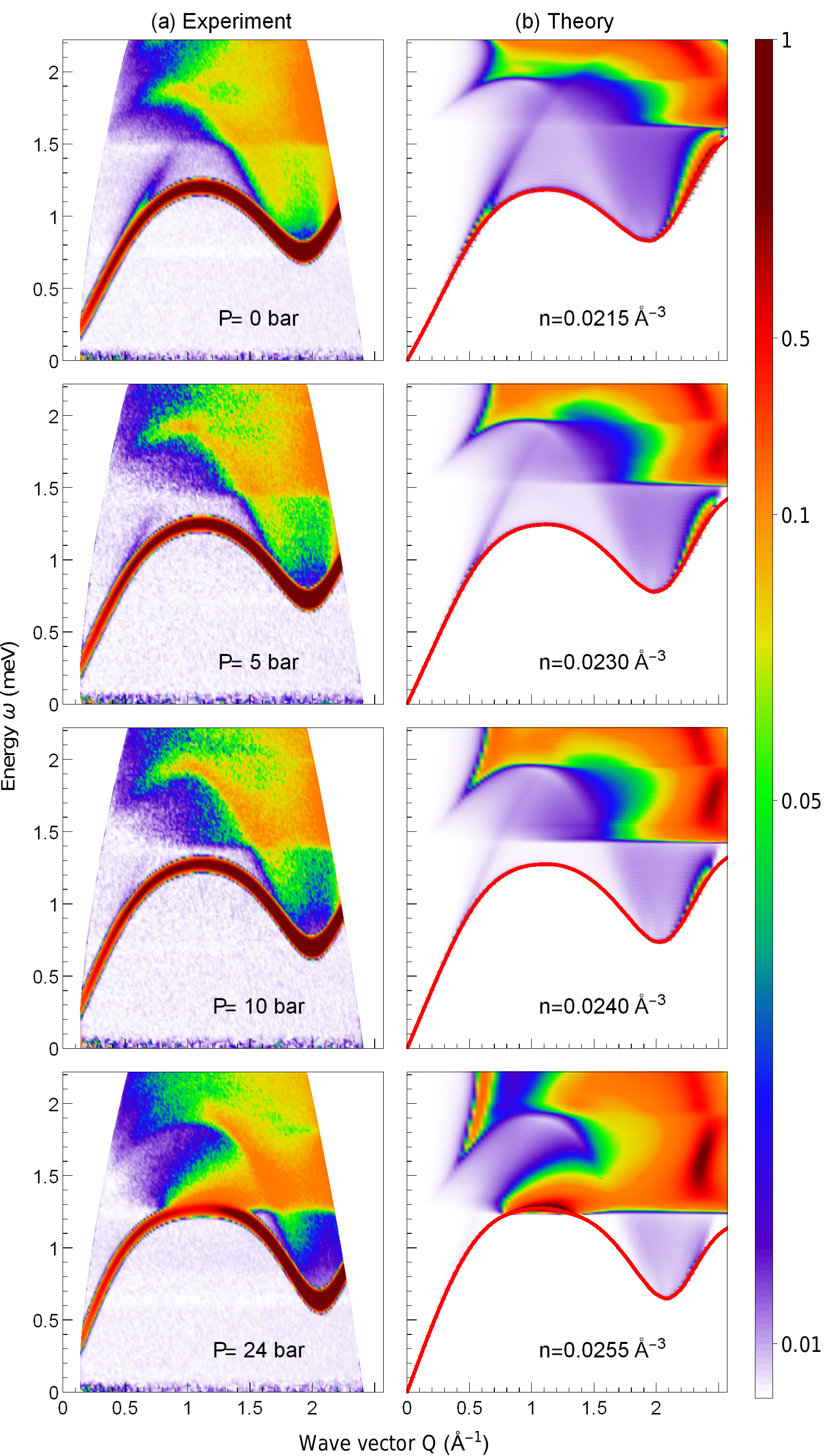}
	\caption{(a)~$S(Q,\omega)$ of superfluid $^4$He measured as a
          function of wave vector and energy transfer, at $P=0$, 5, 10
          and 24 bars and temperature $T\,\le\,100$\,mK.  The incident
          neutron energy is $E_i$=3.55\,meV.  (b)~Dynamic many-body
          theory calculation of $S(Q,\omega)$ at corresponding
          densities ($n$=0.0215, 0.0230, 0.0240 and 0.0255 \AA$^{-3}$,
          see text).  Note that the main features of the experimental
          data are well reproduced.  The color-coded intensity scale
          is in units of meV$^{-1}$.  The intensity is cut off at
          1\,meV$^{-1}$ in order to emphasize the multi-excitations
          region.  The apparent width of the phonon-roton excitations
          in the experimental plot is due to an energy resolution of
          0.07\,meV, while the calculated phonon-roton dispersion
          curve has been highlighted by a thick red line. }
	\label{allpressures}
\end{figure}

We present in this section the spectra obtained using an incident
neutron energy of $E_i$=3.55\,meV, for wave vectors up to
$Q=2.5\,$\AA$^{-1}$ and energies up to $\omega$=2.22\,meV.  The results
are shown in Fig.\,\ref{allpressures}(a), where we represent our
earlier data \cite{Ketty} at SVP, the present data at 5 and 10\,bars,
and the data at $P=24$\,bars discussed in the
previous section.  
One can readily note that both the
single-excitation and the multi-excitation components of the dynamic
structure factor are modified by the pressure.

\begin{figure}[t]
	\includegraphics[width=0.60\columnwidth]{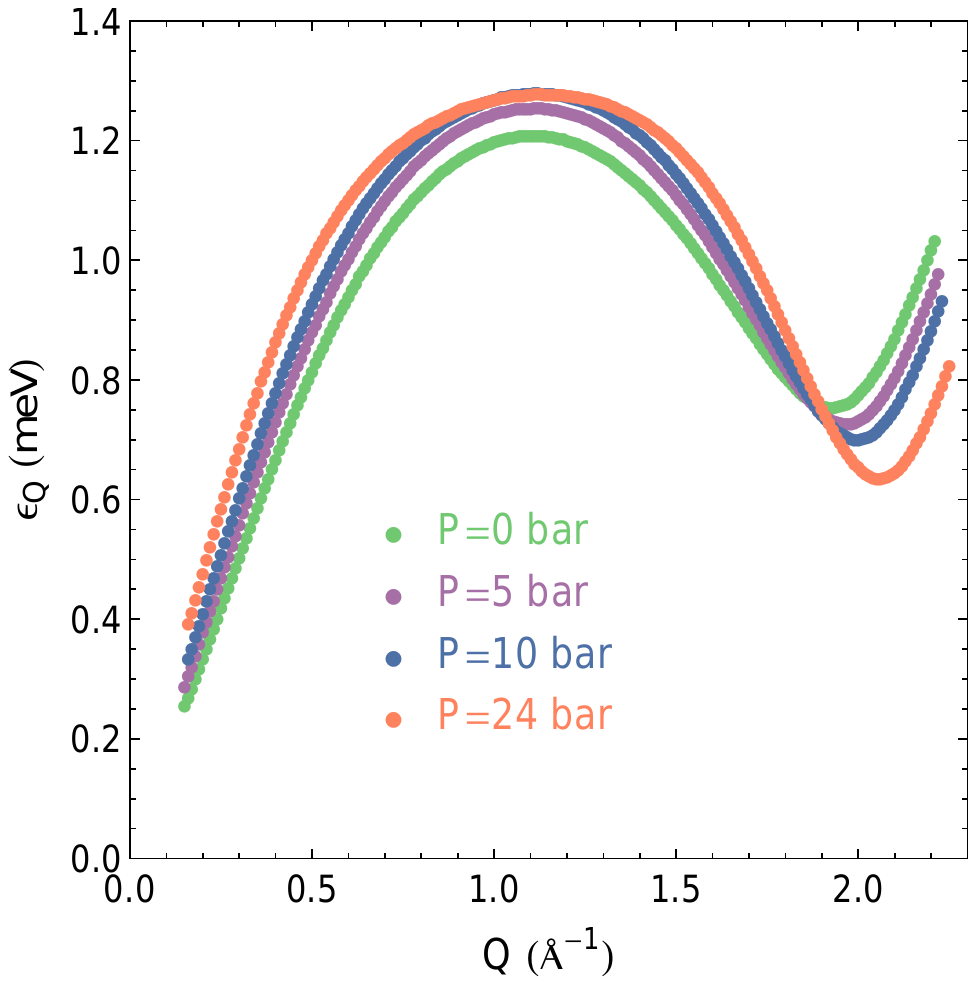}
	\includegraphics[width=0.60\columnwidth]{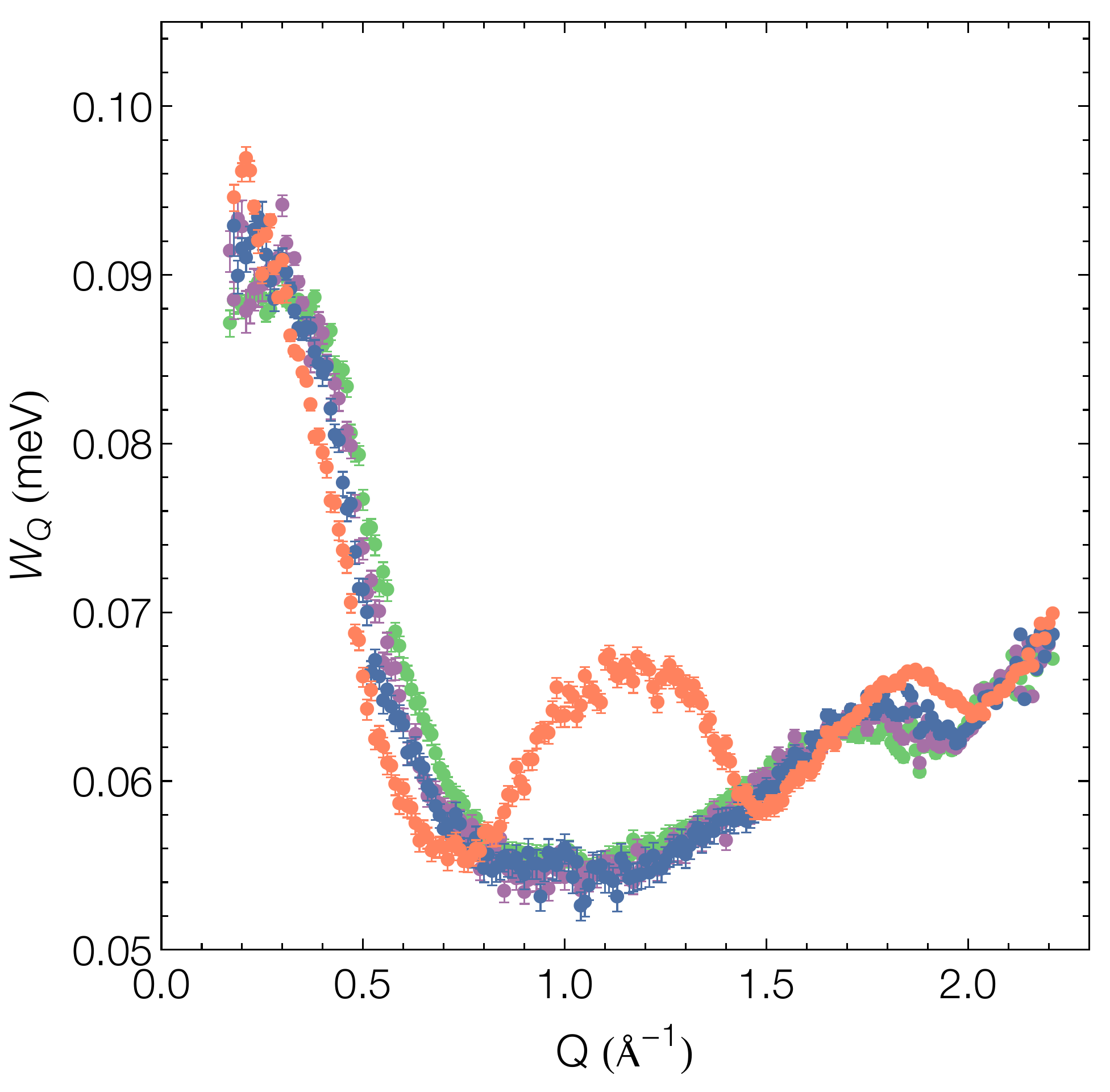}

	\caption{a) Dispersion relation $\epsilon{_Q}(Q)$ of the single-excitations 
          measured at 0, 5, 10 and 24 bars. Note the flattening 
					of the curve at the maxon at high pressures. b) The wave vector
          dependence of the measured width (FWHM) of the
          single-excitation peaks. The measured width reflects the
          shape of the experimental resolution ellipsoid cut by the
          dispersion relation curve at different angles. At 24\,bars,
          however, a physical broadening of the maxon is clearly
          observed.}
		\label{dispersion+width} 
\end{figure}

Our results for the single-excitations dispersion measured at several
pressures, shown in Fig.\,\ref{dispersion+width}a, are in excellent
agreement with previous works
\cite{GlydeBook,Glyde2018review,dietrich72,Woods1977,Stirling91,Andersen92,Gibbs1999,Andersen94a,Keller2004}. 
The roton parameters at each pressure have been obtained from fits of the single-excitations dispersion relation $\epsilon_Q(Q)$ to the expression 
\begin{equation}
	\epsilon_Q=\Delta_R+\frac{\hbar^2}{2m\mu_R}(Q-Q_R)^2+b(Q-Q_R)^3+c(Q-Q_R)^4, 
	\label{Eqfitroton}
\end{equation}
where $\Delta_R$ is the roton energy gap, $Q_R$ the wave vector at the roton minimum, 
and $\mu_R$ the roton effective mass; $b$ and $c$ are additional adjustable parameters. 
Fits were made over a total wave vector range $\Delta$Q up to 0.47\,\AA$^{-1}$. 
Due to the large number of individual detectors and the high neutron rate of IN5, 
the statistical uncertainty of the fits is very good (see Table\,\ref{Tabrotonparameters}). 
The roton mass determined in our work is lower than the one obtained by 
Andersen et al. \cite{Andersen92,Andersen94a} using a parabolic fit of the roton minimum, but it agrees well with 
earlier measurements \cite{Woods1977} where the parabolic fit was limited to a very narrow wave vector range. 

\renewcommand{\arraystretch}{1.3}

\begin{table}[!h]
	\centering
	\begin{tabular}{|c||c|c|c|}
		\hline  P (bars)   & $\Delta_R$ (meV) & $Q_R$ (\AA$^{-1}$) &  $\mu_R$ \\
		\hline  0 & 0.7416(10)   & 1.9260(2)  & 0.1240(4)    \\
		\hline  5.01(2) & 0.7143(10)  & 1.9655(2)  & 0.1096(4)  \\
		\hline  10.01(2) & 0.6885(10)  & 1.9963(2)  & 0.1000(4)   \\
		\hline  24.08(2) & 0.6254(10)  & 2.0579(2)  & 0.0879(4)  \\
		\hline 
	\end{tabular}
	\caption{Roton energy gap $\Delta_R$, wave vector of the roton minimum $Q_R$ and roton effective mass $\mu_R$; values in parenthesis are one standard deviation errors from least-squares fits described in the text.}
	\label{Tabrotonparameters}
\end{table}

A similar analysis can be performed in the maxon region. 
The corresponding maxon parameters $\Delta_M$, $Q_M$ and $\mu_M$ ($d$ and $e$ are additional adjustable parameters) have been calculated by fits of $\epsilon_Q$ in the maxon region, over a wave vector range $\Delta$Q on the order of 0.8\,\AA$^{-1}$, to the formula: 
\begin{equation}
	\epsilon_Q=\Delta_M-\frac{\hbar^2}{2m\mu_M}(Q-Q_M)^2+d(Q-Q_M)^3+e(Q-Q_M)^4.
	\label{Eqmaxonfit}
\end{equation}
The results are given in Table \ref{Tabmaxonparameters}. 

\begin{table}[!h]
	\centering
		\begin{tabular}{|c||c|c|c|c|}
		\hline  P (bars) & $\Delta_M$ (meV)   & $Q_M$ (\AA$^{-1}$) &  $\mu_M$ &  $2\Delta_R$\\
		\hline  0 & 1.1966(10)   & 1.1073(2)  & 0.492(1) & 1.4832(20)   \\
		\hline  5.01(2) & 1.2422(10)   & 1.1089(3)  & 0.541(1) & 1.4286(20) \\
		\hline  10.01(2) & 1.2668(10)  & 1.1150(3)  & 0.614(2) &  1.3777(20) \\
		\hline  24.08(2) & 1.2662(10)  & 1.1336(4)  & 0.915(3) & 1.2508(20)\\
		\hline 
	\end{tabular}
	\caption{Maxon energy $\Delta_M$, wave vector $Q_M$ and effective mass $\mu_M$; values in parenthesis are one standard deviation errors from least-squares fits described in the text. The last column gives twice the energy of
the roton gap $2\Delta_R$, for comparison with the value of $\Delta_M$.}
	\label{Tabmaxonparameters}
\end{table}

As expected for a system approaching localization
\cite{nozieres2004roton}, the phonon and the maxon energies increase
steadily with pressure, while the energy of the roton minimum
decreases. The single-excitation data of
Fig.\,\ref{dispersion+width}a clearly show in addition a substantial
flattening at the level of the maxon in the dispersion curve
corresponding to a pressure of 24\,bars.  Earlier results at this
pressure did not detect this effect \cite{dietrich72,Stirling91},
while more recent systematic results by Gibbs {\em et al.} \cite{Gibbs1999}
were limited to pressures below 20\,bars.  The data at 24\,bars are
qualitatively different from those at low pressures because the maxon
energy exceeds twice the roton energy. At high pressures, the maxon
excitation can therefore decay by phonon emission, exactly as in the case 
of higher wave vectors, at the Pitaevskii's plateau \cite{pitaevskii59}.

We also observe the corresponding broadening of the maxon
single-excitation (Fig.\,\ref{dispersion+width}b): the measured maxon
total width of 0.012\,meV, obtained after subtraction of the
instrumental resolution, is substantial compared to typical phonon and
roton widths (see Ref.\,\onlinecite{Keller2004} and references therein).  
The excitations in the maxon region broaden until they become unobservable 
in confined helium \cite{Glyde2018review,Pearce2004,Bossy2008}, where
very high pressures can be reached before solidification.

We now concentrate on the multi-excitation region, shown in
Fig.\,\ref{allpressures}(a), which displays highly structured spectra
for all pressures.  The data for the pressures 5 and 10 bars are
qualitatively similar to our previous results at saturated vapor
pressure \cite{Ketty}.  The high resolution spectra display very
clearly a threshold in energy at about 1.5\,meV. This feature, which
corresponds to the decay of an excitation into a pair
of rotons, depends on pressure, since the roton energy depends on pressure. 
In addition, we observe several well-defined
multi-excitation branches displaying substantial dispersion. Their
gradual evolution reflects, as will be shown in section
\ref{sec:analysis}, the change with pressure of the single-excitation
dispersion.

We also observe important {\it qualitative\/} changes at high pressures.
We examine first the multi-excitation region of the ``ghost-phonon''.
This multi-phonon excitation, observed in our previous work, 
appears as a linear
extension of the phonon branch \cite{Ketty}.  We observe in the
present work that the ghost-phonon intensity strongly decreases with
pressure until it disappears at some pressure below 24\,bars.

We also see very clearly in Fig.\,\ref{completemap}(a) a multi-phonon
region just above the roton branch for wave vectors of the order of
2.2 to 2.4\,\AA$^{-1}$.  The high resolution spectra at
$E_i$=3.55\,meV only show part of this multi-excitation region, but
the results have been completed by spectra taken at $E_i$= 5.11\,meV
at SVP and 24\,bars, shown in Fig.\,\ref{completemap}.  
The intensity of these multi-excitations, described in detail in 
Section \ref{sec:ghost_roton}, 
decreases strongly with pressure, behaving similarly as the
ghost-phonon.

The multi-excitation spectra are strongly modified at high pressures,
as the maxon enters the multi-excitations continuum.
Fig.\,\ref{allpressures} shows that substantial intensity develops at
this pressure for energies just above the maxon.  
Similar effects were also observed by Graf {\em et al.} \cite{Graf74}, Talbot {\em et al.} \cite{Talbot88}, 
and by Gibbs {\em et al.} \cite{GibbsThesis,Gibbs1999,Gibbs2000} 
at a lower pressure (20\,bars).  The present data benefit from a 
sharper resolution, as can be seen by directly 
comparing spectra at $Q\approx 1$\,\AA$^{-1}$ around the maxon peak.

All these effects will discussed in detail in section
\ref{sec:analysis} in the context of a comparison with theoretical
calculations.

\section{Calculations with the Dynamic Many-body Theory}
\label{sec:theory}

We present in this section our calculations of the dynamic structure
factor of superfluid $^4$He at zero temperature obtained within the
Dynamic Many-Body theory \cite{eomI,eomIII}.

\subsection{State of the art of Theory}

Theoretical studies of the dynamic structure function in $^4$He began
with the work of Feynman \cite{Feynman}, and Feynman and Cohen
\cite{FeynmanBackflow}.  The Feynman theory of elementary excitations
was developed in a systematic Brillouin-Wigner perturbation theory by
Jackson and Feenberg
\cite{RevModPhys.34.686,PhysRev.185.186,PhysRevA.8.1529}.  An
important contribution was the identification of classes of theories
for the dynamic structure function \cite{PhysRevA.9.964} that satisfy
the $\omega^0$ and $\omega^1$ sum rules exactly.

The most complete evaluation of the phonon-roton dispersion relation
in terms of Brillouin-Wigner perturbation theory was carried out by
Lee and Lee \cite{PhysRevB.11.4318} who obtained an impressive
agreement with the experimental phonon-roton spectrum up the wave
vector of 2.5\,\AA$^{-1}$.  The major drawback with these early
calculations was that the required input, pair and three-body
distribution functions, were poorly known.

Manousakis and Pandharipande \cite{Manousakis84,Manousakis86} used
input states of the Brillouin-Wigner perturbation theory including
``backflow'' correlations in the spirit of Feynman and Cohen.  Through
the gradient operator acting on the wave function, specific 
dynamic correlations are introduced to all orders. The
``backflow-function'' is, however, chosen per physical intuition rather
than by fundamental principles, and the evaluation of the perturbative
series becomes very complicated. Topologically, diagrams similar to
those of Lee and Lee \cite{PhysRevB.11.4318} were included.  While the
accuracy of the theoretical roton energy is comparable to that of Lee
and Lee, one can clearly see an inconsistency since the energy of the
Pitaevskii plateau \cite{pitaevskii59} lies below twice the energy of
the roton gap.

The first theoretical descriptions of the multi-excitations
\cite{Gotze76,Manousakis84,Manousakis86} were qualitatively in
agreement with the early multi-excitations data
\cite{CowleyWoods,dietrich72,Graf74}. The simplest version of Correlated
Basis Functions theory produces phonon, maxon and roton modes, as well
as multi-phonons. 
In this approximation, the calculated multi-excitations decay into
Feynman modes instead of the correct single-excitations; large
gaps are found in the spectrum, and many predicted features are not
seen in the experiments.  Other features calculated in the
multi-excitation region do indeed survive in recent theories, like the
presence of intensity above the phonon branch and that of a
well-defined 2-roton threshold (these effects are described below).
These calculations, as well as many others addressing specific aspects
of the multi-excitation dynamics, could not be quantitatively compared
to the experimental results, but they motivated further investigations
on multi-particle dynamics. Reviews can be found in
Ref.\,\onlinecite{GlydeBook,TriesteBook}.

More recent calculations \cite{eomIII} used a hybrid approach of
Brillouin-Wigner perturbation theory and equations of motion for
time-dependent multi-particle correlation functions to derive a
self-consistent theory of the dynamic density-density response of
$^4$He.  The self-consistency of this semi-analytic method allows the
identification of mode-mode coupling processes that lead to observable
features in the dynamic structure function.  The underlying physical
mechanisms, their relationship to the ground state structure, and the
consequences on the analytic properties of the dynamic structure
function, emerge directly from the theory.

A very different approach involves novel numerical methods
\cite{Ceperley1996,Vitali,roggero2013dynamical,nava2013dynamic,ferre2016dynamic}
that give access to dynamic properties of quantum fluids. These
important algorithmic developments will reproduce, extend and complete  
the experimental data with the future development of computing power;
their present accuracy and consistency, however, are still limited in
the multi-excitations region investigated here.

\subsection{Dynamic Many-Body Theory calculation}

In order to calculate quantitatively both the single-excitation and
the multi-excitation response, our calculations include up to
three-body dynamic fluctuations in the correlation functions of the
equations of motion \cite{eomIII}.  We derive the self-consistent
density-density response of $^4$He $\chi$(Q,$\omega$), expressed as

\begin{equation}
	\chi(Q,\omega)=\frac{S(Q)}{\omega-\Sigma(Q,\omega)}+\frac{S(Q)}{-\omega-\Sigma(Q,-\omega)}
\end{equation}
where $S(Q)$ is the static structure factor, and the self-energy
$\Sigma(Q,\omega)$ is determined by the integral equation

\begin{multline}
%\begin{equation}
	\Sigma(Q,\omega)=\epsilon_0(Q)+ 
	\frac{1}{2}\int\frac{d^3pd^3k}{(2\pi)^3n}
        \delta(\vec{Q}-\vec{p}-\vec{k}) \times \\
	\frac{|V^{(3)}(\vec{Q};\vec{p},\vec{k})|^2}
             {\omega-\Sigma(p,\omega-\epsilon_0(k))
               -\Sigma(k,\omega-\epsilon_0(p))}\,.
	\label{eqTheory}
%\end{equation}
\end{multline}
In this expression, $\epsilon_0(Q)$ is the Feynman dispersion relation,
and $V^{(3)}$ the three-body coupling matrix element. The simplest
approximation for $V^{(3)}$, the so-called convolution approximation
\cite{PhysRevA.8.1529}, including static ground state triplet
correlations \cite{Chuckphonon}, improves the density--dependence of
the roton minimum visibly. The most advanced calculation \cite{eomI},
which is taken here and in Ref.\,\onlinecite{eomIII}, sums an infinite
series of diagrams, the so-called ``fan-diagrams'' which is the
minimum set of diagrams that must be included to reproduce exact
features of $V^{(3)}$ for both, long wavelength and short distances.

Linear response theory \cite{GlydeBook,eomIII} provides the
relation between the experimental dynamic structure factor and the
dynamic susceptibility calculated by the theory described above: the
dynamic structure factor $S(Q,\omega)$ is proportional to the
imaginary part of the dynamic susceptibility $\chi(Q,\omega)$, the
linear response of the system to a density fluctuation.

Full maps of $S(Q,\omega)$ have been calculated for different atomic
densities, see Fig.\,10 in Ref.\,\onlinecite{eomIII}. The data
shown in Figs.\,\ref{completemap} and \ref{allpressures} correspond to
$n=0.0215$, 0.0230, 0.0240 and 0.0255 \AA$^{-3}$, values which provide
the best overall agreement with the experiment.  They turn out to be
very close to the experimental results for $P=0$, 5, 10 and 24\,bars,
$n_{\rm exp}$=0.0218, 0.0230, 0.0239 and 0.0258 \AA$^{-3}$. The small
shift in density is within the expected accuracy of the theoretical
calculations.

The calculations presented here have been performed using only the
most relevant diagrams \cite{eomIII}. This approximation is sufficient
to provide an excellent description of the dynamics, but minor
discrepancies can still be seen. The most salient effect is that the
roton energy is overestimated; at zero pressure, for instance, the
calculated value is 0.83\,meV while the measured value is
0.7416(10)\,meV. This discrepancy could be resolved by including
additional diagrams, but it does not seem necessary to perform such a
tedious calculation given the quality of the agreement already
achieved at this stage.

The calculation provides absolute values for the structure factor. 
In our previous work \cite{Ketty}, the calculated 
values were multiplied by an
overall normalization factor of 1.28 in order to have
$Z(Q=2\,$\AA$^{-1}$)=0.93 near the roton. 
Here, this normalization has not been applied. 
Given the finite number of diagrams involved in the calculations, 
a factor of this order is within their estimated absolute accuracy.

\subsection{Mode-mode coupling}
Multi-excitations arise from the enhanced response of the system at
particular energies and wave vectors corresponding to two or more
single-excitations into which they can decay.  The theory considers
(see equation \ref{eqTheory}) the most relevant processes where a
density fluctuation ($\vec{Q},\omega$) of wave vector $\vec{Q}$ and
energy $\omega$ decays into a pair of single-excitations with
corresponding values ($\vec{p},\omega_p$) and ($\vec{k},\omega_k$).
The calculations have been shown to be in excellent agreement with
experiment at saturated vapor pressure \cite{Ketty}.  Here we
investigate the general pressure dependence of the dynamics, and
several particularly intense mode-mode couplings. The latter were
examined theoretically in Ref.\,\onlinecite{eomIII}, and additional
calculations specialized to the main mode-mode couplings
(phonon-phonon, phonon-roton, roton-roton, maxon-roton) can be found
in Ref.\,\onlinecite{modemode}.  The next section provides a detailed
comparison between the theory and the experimental data.

\section{Identification of the multi-excitations}
\label{sec:analysis}

Above the sharp and intense phonon-maxon-roton dispersion curve, we
observe a highly-structured multi-excitation region.
Multi-excitations are relatively strong if they can decay into a pair
of high intensity single-excitation modes. The energy and momentum of
these pair combinations is directly related, by the conservation of
energy and momentum, to those of the underlying elementary
excitations. It is possible to determine  the position of
the main multi-excitation resonances in the dynamic structure factor
map (2-Phonons, 2-Rotons, 2-Maxons and Maxon-Roton) from pure kinematic
considerations, {\em i.e.\/} energy and momentum conservation. 
The challenge for
microscopic theories is to predict the {\it{intensity\/}} of the
multi-excitation spectrum, if possible in a large dynamic
range. Obtaining the fine structure we observe requires a quantitative
calculation of mode couplings.

We first present in this Section a brief description of the kinematic 
constraints for different pair-excitations, setting the framework 
for their identification. 
The following two subsections concentrate on 
new features observed in the multi-excitation spectrum, that we named 
``ghost-phonon'' and ``ghost-roton''.  
We then describe a different type of multi-excitations, associated to 
roton-roton coupling, which we  
observed in particular ``above the maxon'' and
``beyond the roton''. 
We conclude this Section by a discussion on higher order multi-excitations, 
and the progressive evolution to the high energy regime.

\subsection{Kinematic constraints for pair-excitations}
\begin{figure}[h]
	\includegraphics[width=0.9\columnwidth]{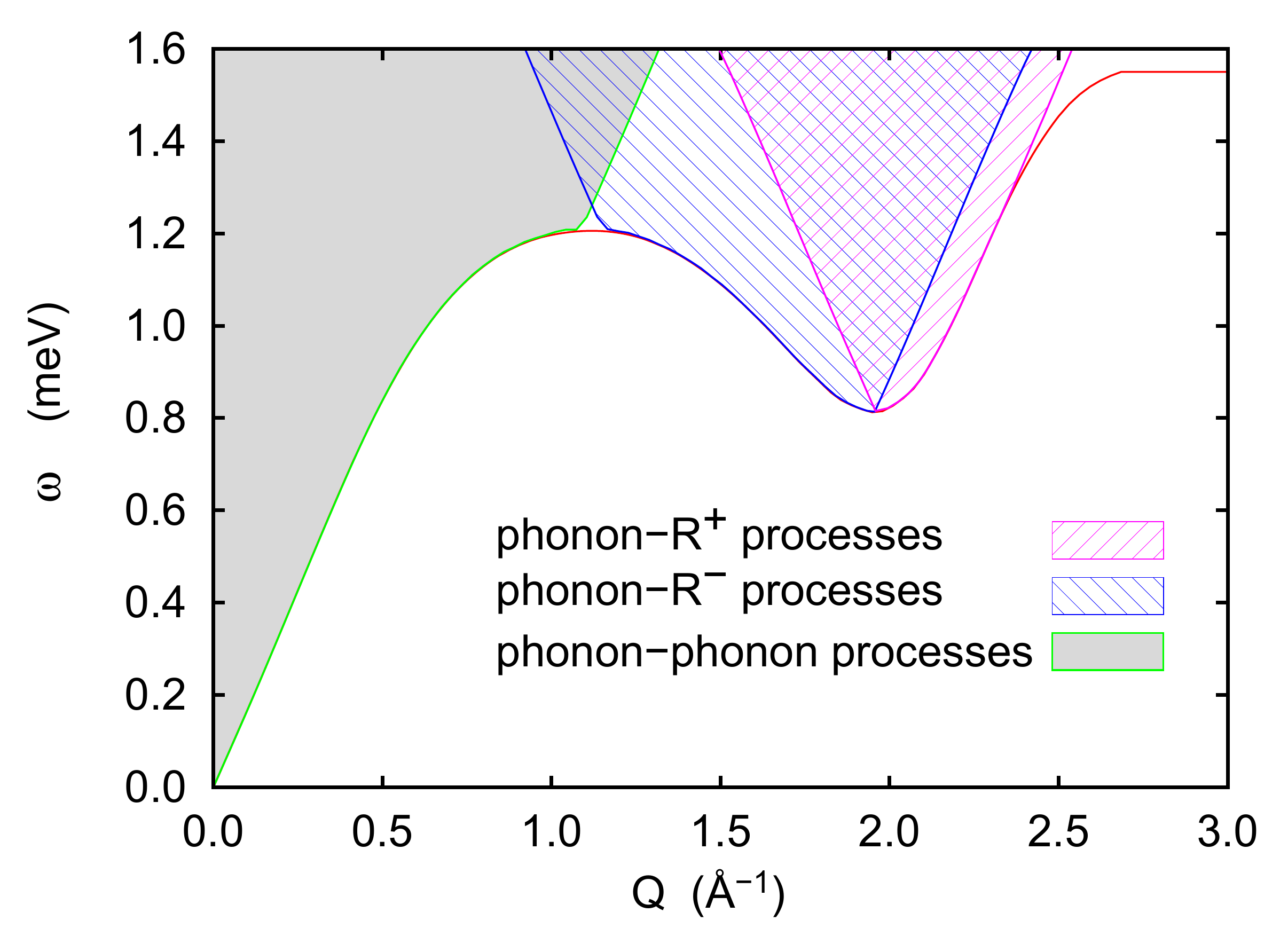}  
	\caption{Kinematically allowed regions for different multi-phonon processes: P-P (including P-M$^-$), P-R$^-$, and P-R$^+$. See text for details.} 
	\label{kinematics}
\end{figure}

The kinematic constraints calculated for the main low energy
multi-excitations are shown in Fig.\,\ref{kinematics}.  We use below
the notation R$^-$ and R$^+$ to distinguish rotons on each side of the
roton minimum.
%, and M$^+$ and M$^-$ is the analogue for maxons.

The allowed regions are necessarily located above the
single-excitations dispersion curve.  The P-P region is found at low 
wave vectors.  Beyond the maxon, P-R$^-$ excitations are allowed in a
large region delimited by the dispersion curve and two lines starting
at the maxon maximum and at the roton minimum, with slopes equal to
$-c$ and $+c$, respectively, where $c$ is the speed of sound.  P-R$^+$
excitations occupy a region delimited by the dispersion curve and a
line starting from the roton minimum with slope $-c$.  There is a
large overlap with the P-R$^-$ region.  

\begin{figure}[t!]
	\includegraphics[width=0.95\columnwidth]{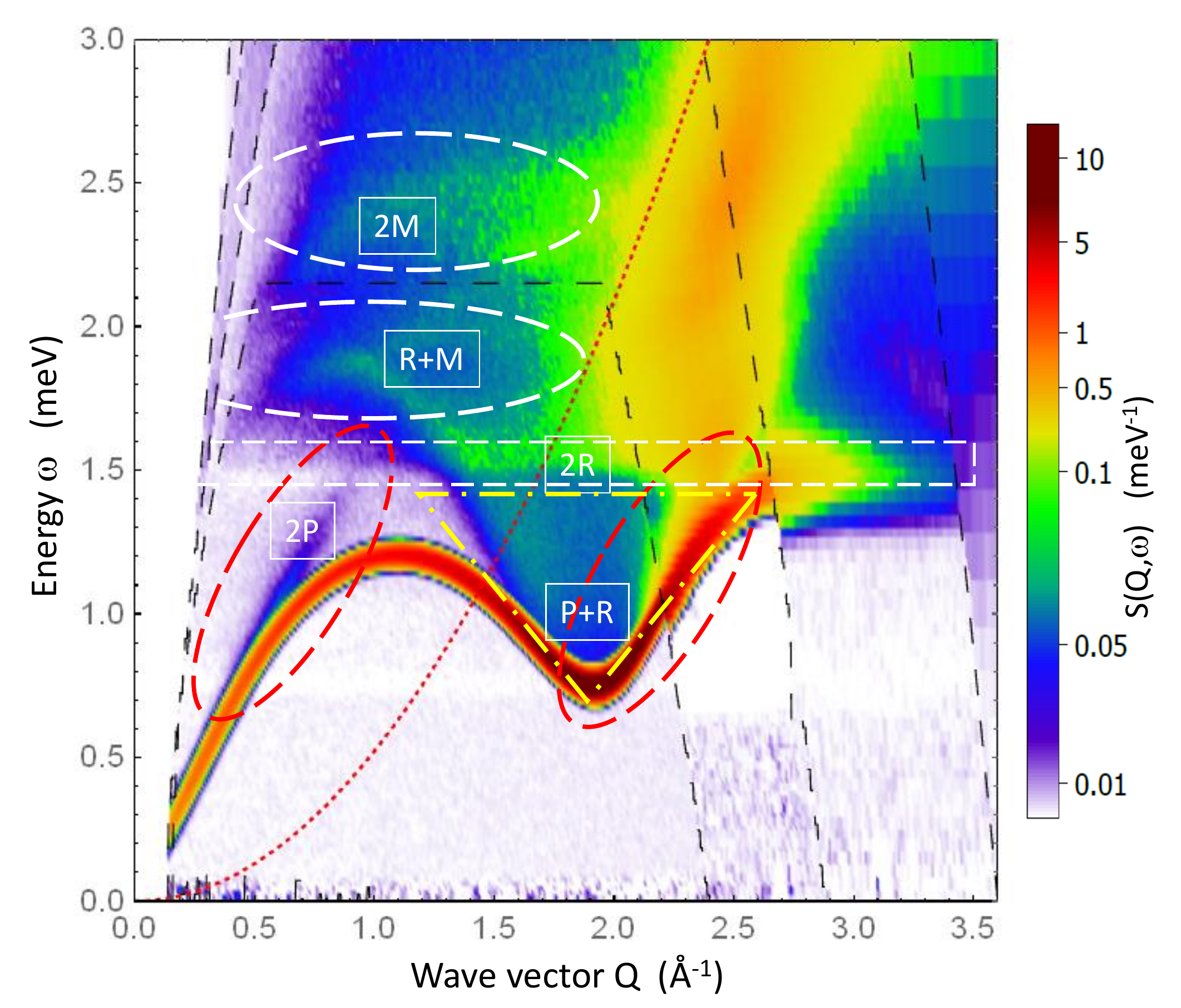}
	\caption{Map of $S(Q,\omega)$ at SVP identifying remarkable mode-mode
          coupling regions: phonon-phonon (2P, with an ellipse
          around the ``ghost-phonon''),
          phonon-roton (P+R, a region marked by a triangle, which includes an ellipse
          indicating more specifically a high-intensity ``ghost-roton'' region), 
					roton-roton (2R, marked by a rectangle 
          around 1.5\,meV), and at higher energies the
          roton-maxon (R+M) and maxon-maxon (2M) regions. The description of the
          different lines is given in Fig.\,\ref{completemap}.  }
	\label{identification}
\end{figure}

The case of 2R, not shown, is
particularly simple, with a threshold at twice the roton energy,
$2\Delta{_{R}}$.  The situation for 2M processes is similar, with an
upper limit equal to 2$\Delta{_{M}}$.  M-R combinations of excitations
may lead to branches with substantial dispersion.  The kinematic
constraints are sufficient to determine unambiguously which are the
dominant processes in some multi-excitations regions, in particular at
low $Q$ above the phonon dispersion, and inside the roton parabolic
dispersion curve. 

The evolution of the observed multi-excitations in a large energy
range, for different pressures, is illustrated in
Figs.\,\ref{completemap} and \ref{allpressures}.  We can distinguish
different types of multi-excitations.  Several narrow branches are
easily identified, as indicated in Fig.\,\ref{identification}, as
corresponding to 2P, 2R, 2M and M-R processes. The 2R feature is
observed in Fig. \ref{allpressures} as a clear threshold, both in the
theoretical and experimental data.  

\subsection{Phonon-phonon coupling: the ghost-phonon}

\begin{figure}[h!]
\includegraphics[width=1.0\columnwidth]{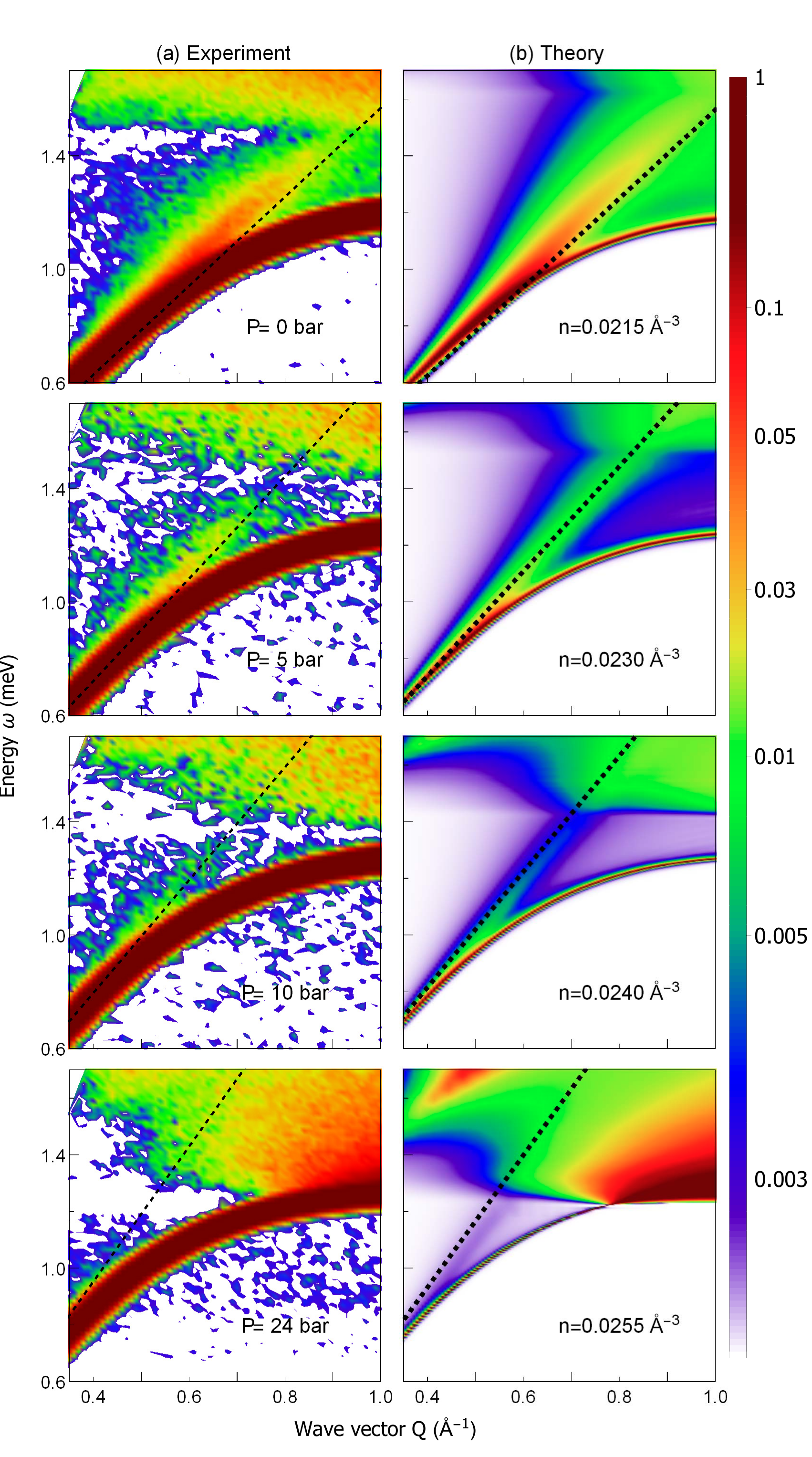}
	\caption{Left: measured dynamic structure factor $S(Q,\omega)$
          in the ghost-phonon region at $P=0$, 5, 10 and 24\,bars. The
          dashed straight lines correspond to the sound dispersion curve at each pressure,
          taken from direct measurements of the sound velocity
          \cite{abraham70}. Right: calculated dynamic structure factor
          at corresponding densities, $n=0.0215$, 0.0230, 0.0240 and
          0.0255\,\AA$^{-3}$, respectively (see text). The dashed straight lines correspond
          to the calculated sound velocities. The color-coded
          intensity scale is in units of meV$^{-1}$.}
		\label{ghost-phonon-2Dplots}
\end{figure}

The ghost-phonon \cite{eomIII,Ketty} (see Section
\ref{subsec:effect-of-pressure} and Fig. \ref{identification})
corresponds to a process where a high energy
multi-excitation decays into a pair of phonons of lower energy. 
In the case of phonon {\em single-excitations}, anomalous
dispersion opens the phase space needed for such processes.  The
anomalous character of the phonon dispersion strongly decreases with
increasing pressure, and normal dispersion is recovered at high
pressures \cite{Maris1970,MarisRMP,GlydeBook,Glyde2018review}.  The ghost-phonon
intensity follows this trend: the pressure dependence is strong, and
the ghost-phonon is clearly suppressed at $P=24$\,bars, as shown in
the experimental and theoretical results in Fig. \ref{allpressures},
and in more detail in Fig. \ref{ghost-phonon-2Dplots}.

\begin{figure}[h]
	\includegraphics[width=1.0\columnwidth]{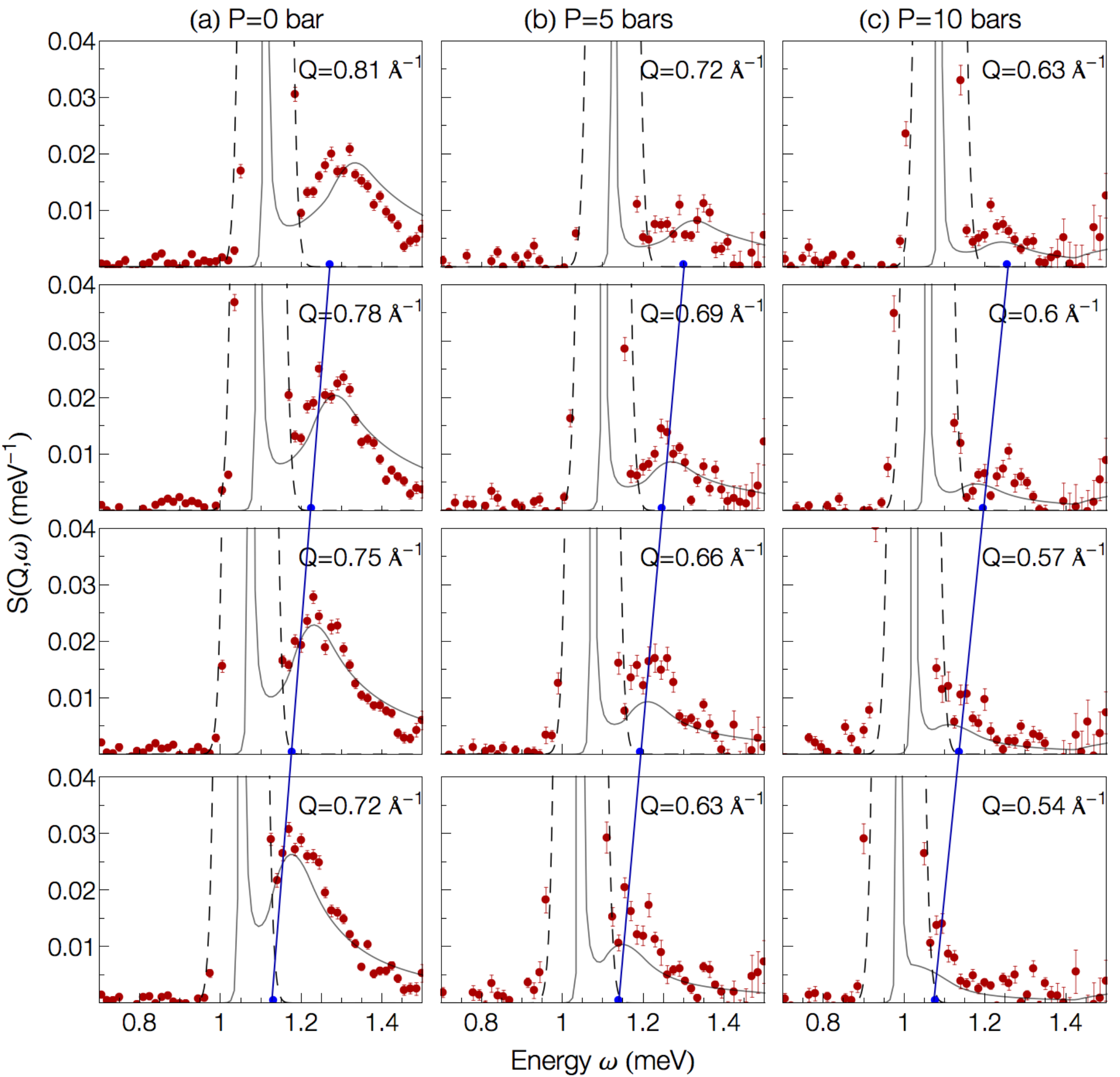}
	\caption{Dynamic structure factor $S(Q,\omega)$ in the
          ghost-phonon region: spectra for different wave vectors $Q$ at
          (a) $P=0$\,bar, (b) $P=5$\,bars and (c) $P=10$\,bars.  Filled
          circles: Experimental $S(Q,\omega)$.  Theoretical dynamic
          structure factor spectra shown as solid lines at densities
          $n=0.0215$, 0.0230 and 0.0240 \AA$^{-3}$. Dashed lines:
          Intensity of the phonon-roton mode (cut off) calculated
          directly from the self energy \cite{eomIII} and convolved
          with the instrumental resolution of 0.07\,meV.  The blue
          lines represent the linear phonon dispersion
          $\epsilon_Q(P)/\hbar Q=C_0(P)$, where $C_0(P)$ is the sound
          velocity at a given pressure \cite{abraham70}.}
		\label{ghost-phonon}
\end{figure}

Cuts of $S(Q,\omega)$ at several wave vectors at the ghost-phonon
level are presented for $P$=0, 5 and 10\,bars in
Fig.\,\ref{ghost-phonon}.  The ghost-phonon peaks for the different
wave vectors are clearly located on the extension of the linear part
of the phonon branch.  According to the calculations [see Eq. (6.4) of Ref.  \onlinecite{eomIII}],
the ghost-phonon remains visible until twice the wave vector up to
which the dispersion relation is to a good approximation linear. 
Indeed, Fig.\,\ref{ghost-phonon} shows that the energy, strength and
shape of the calculated ghost-phonon are in excellent quantitative
agreement with the experiment at all pressures.

\subsection{Phonon-roton coupling and the emergence of the ghost-roton }
\label{sec:ghost_roton}

\begin{figure}[t!]
	\includegraphics[width=0.70\columnwidth]{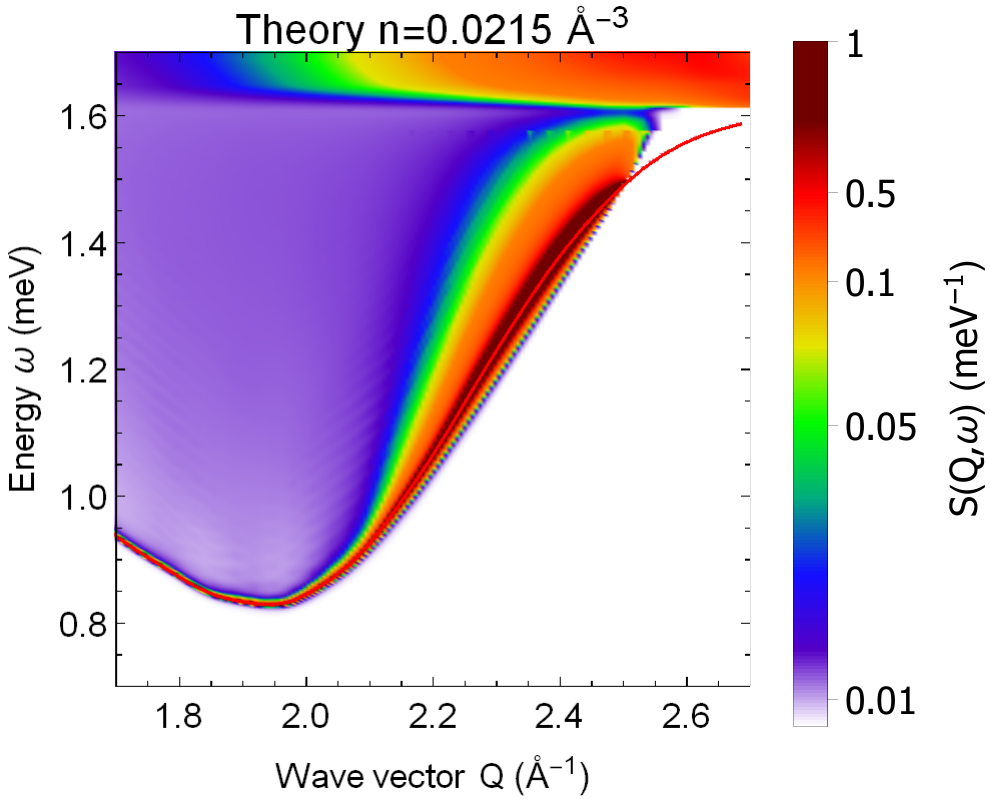}\\
    \vspace{0.5cm}
	\includegraphics[width=0.70\columnwidth]{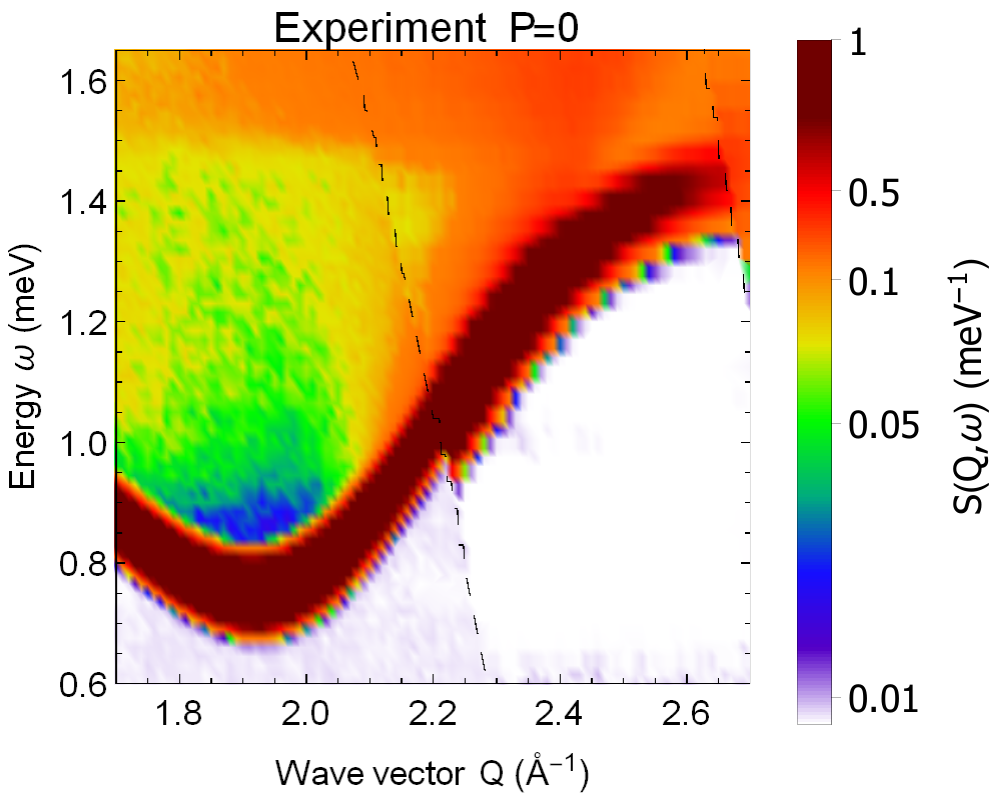}
	\caption{Theoretical and experimental results for
          $S(Q,\omega)$ at saturated vapor pressure displaying
          enhanced multi-excitations (``ghost-rotons'') above the R$^+$
          roton branch, in the supersonic rotons region. The dashed
          lines represent the limits of different neutron kinetic
          ranges, see Fig.\,\ref{completemap}. 
					The small oscillations observed along some contours should be 
					disregarded, they result from numerical discretization.}
  \label{ghost-roton-P0}
\end{figure}

One notes in Fig. \ref{allpressures}, for all pressures, the presence
of substantial intensity in the region within the roton parabola. 
Near the roton minimum, where P-R processes are expected
to dominate, we observe that the intensity is not symmetric with
respect to the roton minimum wave vector $Q_{R}$: a faint branch,
clearly related to the kinematic limitation for P-R$^+$ processes (see
Fig.\,\ref{kinematics}) is seen for $Q\,<\,Q_{R}$, while a strong
branch is formed just above the
dispersion curve for $Q>Q{_{R}}$.  
These new features, and in particular the one for $Q>Q{_{R}}$, provide   
a significant contribution to
the multi-excitations weight at low pressures
(Fig.\,\ref{ghost-roton-P0}). 
They appear as an extension of the roton
parabolic dispersion towards higher energies, and by analogy with
the ghost-phonon, we call these multi-excitations ``ghost-rotons''.

\begin{figure}[h!]
  \includegraphics[width=0.70\columnwidth]{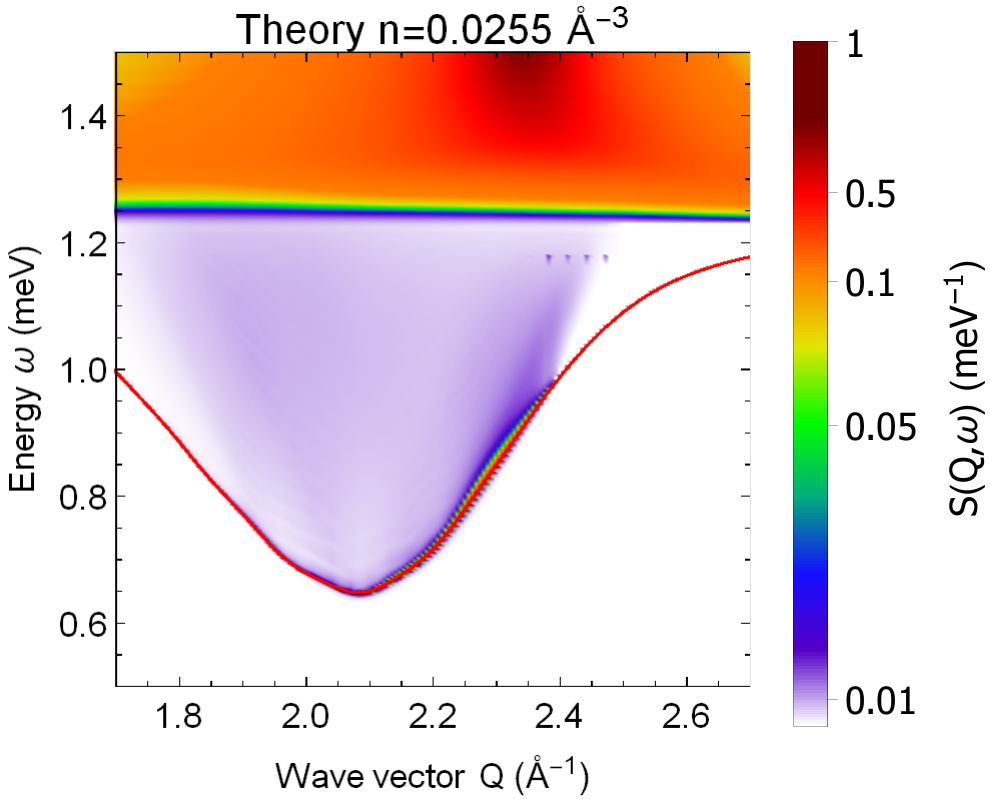}\\
    \vspace{0.5cm}
  \includegraphics[width=0.70\columnwidth]{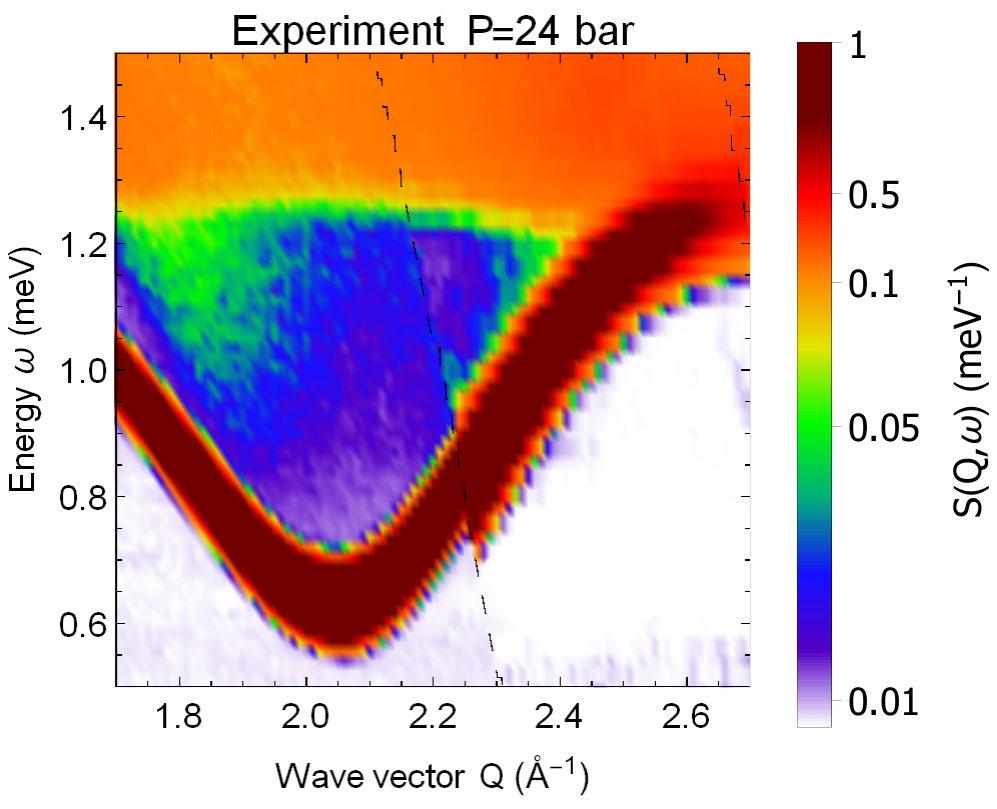}
 \caption{Theoretical and experimental results for $S(Q,\omega)$ at
   $P=24$\,bars. A comparison with Fig.\,\,\ref{ghost-roton-P0} shows
   that at high pressures, ghost-roton multi-excitations are strongly 
   suppressed. They are masked by the finite energy resolution in the experimental graph, but still visible in the calculation.}
	\label{ghost-roton-P24}
\end{figure}

\begin{figure}[t!]
	\includegraphics[width=0.80\columnwidth]{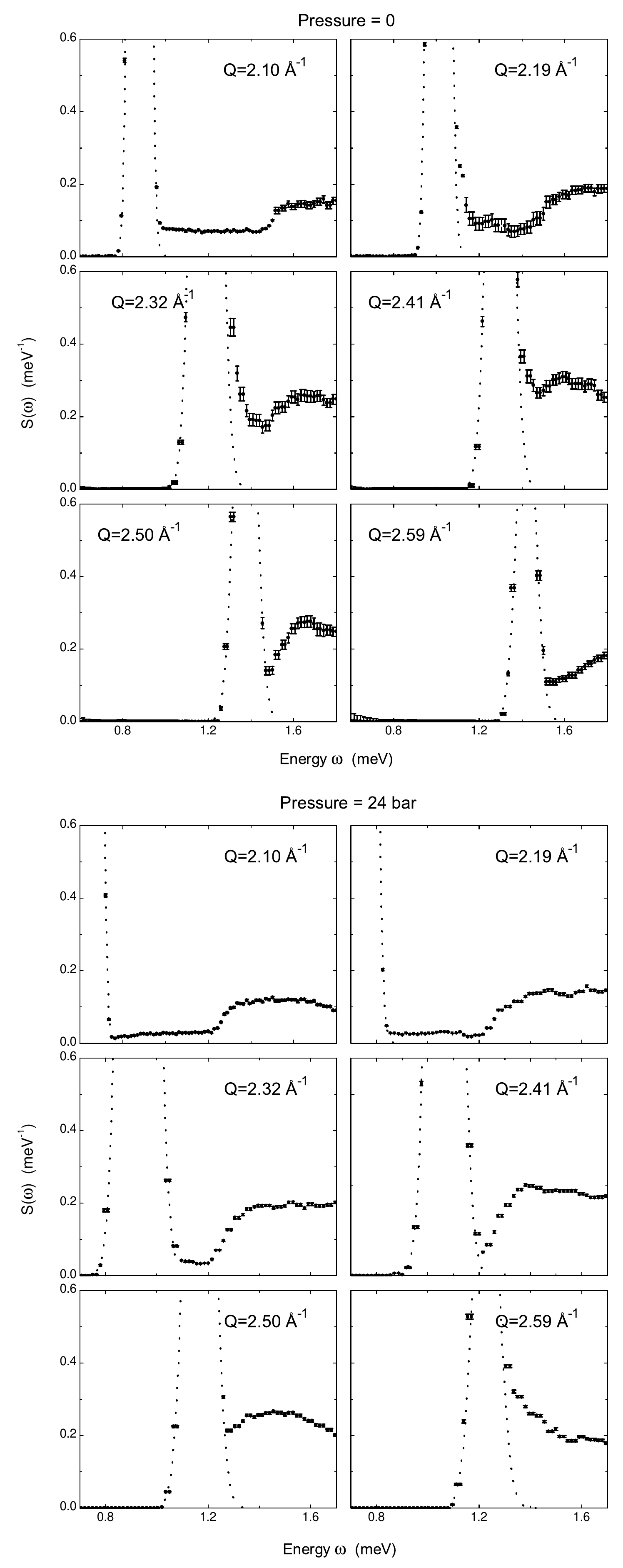}
	\caption{$S(Q,\omega)$ measured in the region of the ghost-roton  
	at $P=0$ (upper graph). At $P=24$\,bars (lower graph) one observes    
	the suppression of the ghost-roton. Dashed lines are Gaussian fits of the single-excitation peaks. 
	A comparison with Figs.\,\ref{ghost-roton-P0} and \ref{ghost-roton-P24} 
	clarifies the origin of the observed roton-peak asymmetry for some wave vectors.}
		\label{cuts-ghost-roton} 
\end{figure}

It is remarkable that the intensity in this region of the P-R
multi-excitations, as was the case for the ghost-phonon, is high at
$P=0$, but is suppressed in the 24\,bars data, as shown in
Figs.\,\ref{ghost-roton-P0},
\ref{ghost-roton-P24} and
\ref{cuts-ghost-roton}. The origin of these effects 
is discussed below.

\begin{figure}[t!]
	\includegraphics[width=0.90\columnwidth]{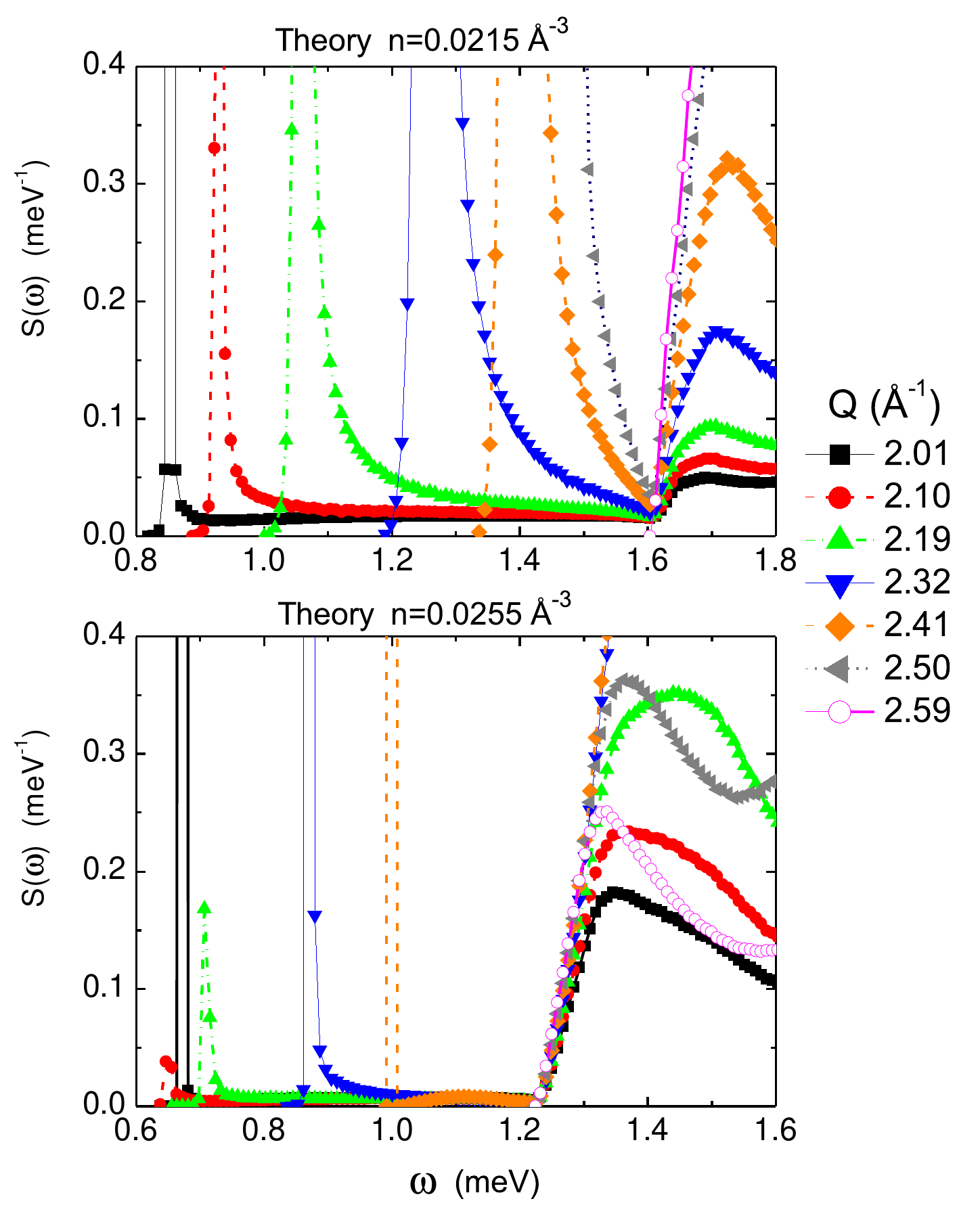}
	\caption{$S(Q,\omega)$ calculated for wave vectors in the
          region of the ghost-roton, at densities $n=0.0215$ and 0.0255\,
          \AA$^{-3}$, associated to $P=0$ and $P=24$ bars respectively.}
		\label{cuts-ghost-roton-theory} 
\end{figure}

Spectra for several wave vectors in the region of the ghost-roton are
shown in Fig.\,\ref{cuts-ghost-roton} at $P=0$ and 24\,bars 
(experiment), and in
Fig.\,\ref{cuts-ghost-roton-theory} for the corresponding densities
$n=0.0215$ and 0.0255 \AA$^{-3}$ (theory). We observe a good 
agreement between theory and experiment, even  
at the highest densities, near solidification. Studies 
of mode-mode couplings \cite{modemode,Fak12} can therefore be
most conveniently performed in the ghost phonon and the ghost-roton
regions, rather than looking for a very small broadening of single
excitations.

%\begin{figure}[h!]
%	\includegraphics[width=0.90\columnwidth]{Fig13-Cuts-Ghost-Roton-P24.pdf}
%	\caption{$S(Q,\omega)$ measured for $P=24$\,bars at wave
%         vectors in the region of the suppressed ghost-roton. 
%					Lines are guides to the eye. See
%          also Fig.\,\ref{ghost-roton-P24}.}
%		\label{cuts-ghost-roton-P24} 
%\end{figure}

%\subsection{Phonon-Cherenkov multi-excitation mechanism}
%\label{subsec:Cherenkov}

Pitaevskii \cite{pitaevskii59} 
described the decay of {\it{single-excitations\/}} when their group
velocity reaches the velocity of sound. He named 
this mechanism of single-excitation broadening "`type a"'. 
The process considered here,
however, is the emission of phonons by {\it{multi-excitations\/}} in the
vicinity of nearly supersonic single-excitations.  The generation of
multi-excitations by neutron scattering in the R$^+$ rotons region by
this mechanism was qualitatively predicted by Burkova
\cite{Burkova1981}.  Here we show that the ghost-roton corresponds to
this effect, that the ghost-phonon is a similar effect, involving
supersonic phonons, and that both are correctly predicted by the
Dynamic Many-Body Theory \cite{eomIII}.
 
\begin{figure}[h!]
	\includegraphics[width=0.70\columnwidth]{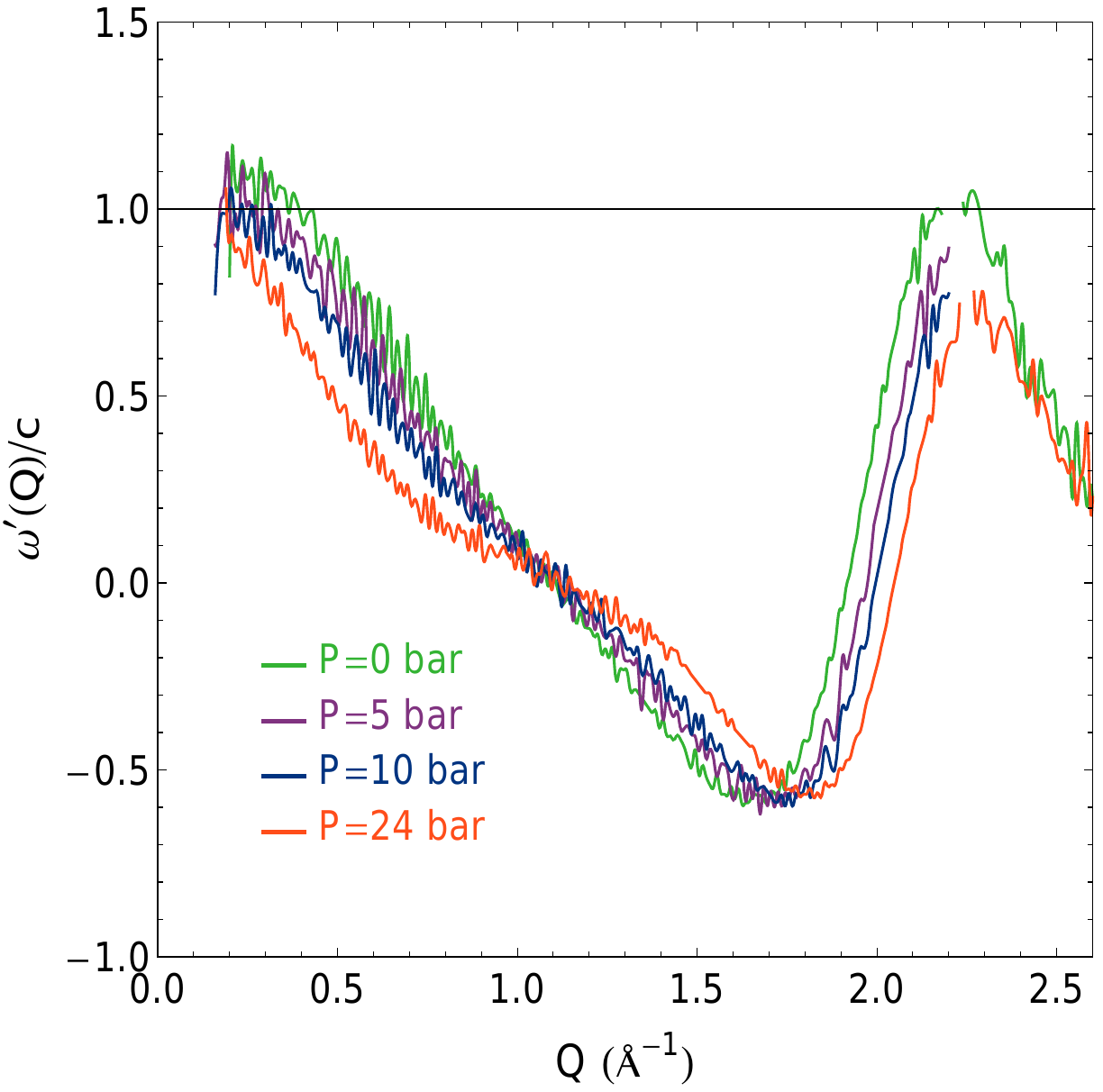}
	\caption{The experimentally determined group velocity of
          single-excitations normalized by the sound velocity
          \cite{abraham70}, as a function of wave vector for
          several pressures.}
	\label{GroupVelocity}
\end{figure}

\begin{figure}[h!]
	\includegraphics[width=0.8\columnwidth]{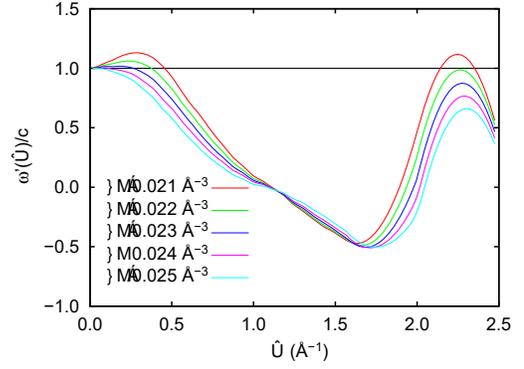}
	\caption{The theoretically calculated group velocity of
          single-excitations normalized by the calculated sound velocity,
          as a function of the wave vector, for
          several densities. The experimental values of the densities
          for $P=0$, 5, 10 and 24\,bars are $n_{exp}$=0.0218, 0.0230,
          0.0239 and 0.0258 \AA$^{-3}$.}
 	\label{soundvelocity}
\end{figure}

It has been observed by Dietrich {\em et al.}\cite {dietrich72} and
confirmed by several groups (see \cite {Keller2004} and references
therein) that the R$^+$ rotons group velocity reaches the sound
velocity for $Q \approx$ 2.2\,\AA$^{-1}$ at low pressures, but remains
below the speed of sound near the melting pressure.  We show in
Figs.\,\ref{GroupVelocity} and \ref{soundvelocity} our measured and
calculated curves for the group velocity of the single-excitations,
for different pressures.  Two regions of interest are clearly seen:
the first one, at low wave vectors, corresponds to the anomalous
dispersion region and gives rise to the ghost-phonon, while the second
occurs for wave vectors somewhat above that of the roton minimum 
(and slightly below the roton minimum, but with a much smaller intensity),
producing the ghost-roton.

According to the analytic calculations by Burkova \cite{Burkova1981},
the neutron-scattering spectrum which corresponds to the production of
one roton should have a linear wing on the high-energy side, with a
slope which depends on the wave vector. This is not really observed,
neither in the experimental data, nor in the Dynamic Many-Body
calculation: the linear part, if any, is probably not visible at the
scale of the graphs (see Figs.\,\ref{ghost-roton-P0},
\ref{ghost-roton-P24}, 
\ref{cuts-ghost-roton} and \ref{cuts-ghost-roton-theory}), or is
buried inside a broadened single-excitations branch.

Several effects are thus observed when the roton single-excitations
approach the speed of sound: a broadening of the roton branch, a
downward bending of the dispersion curve, and the appearance of a
multi-phonon region just above the distorted dispersion curve. These
effects are large at low pressures; the rapid increase of the sound
velocity with pressure is responsible for the suppression of the
ghost-roton multi-excitations at 24 bars.

\subsection{Roton-roton coupling}

We discuss now a different type of {\it{multi-excitations\/}}, related
to Pitaevskii's ``type b'' {\it{single-excitations\/}} decay processes
where the disintegration of a single-excitation occurs as its energy
exceeds twice the roton gap \cite{pitaevskii59,Burkova1981}.

\begin{figure}[h]
  \includegraphics[width=0.8\columnwidth]{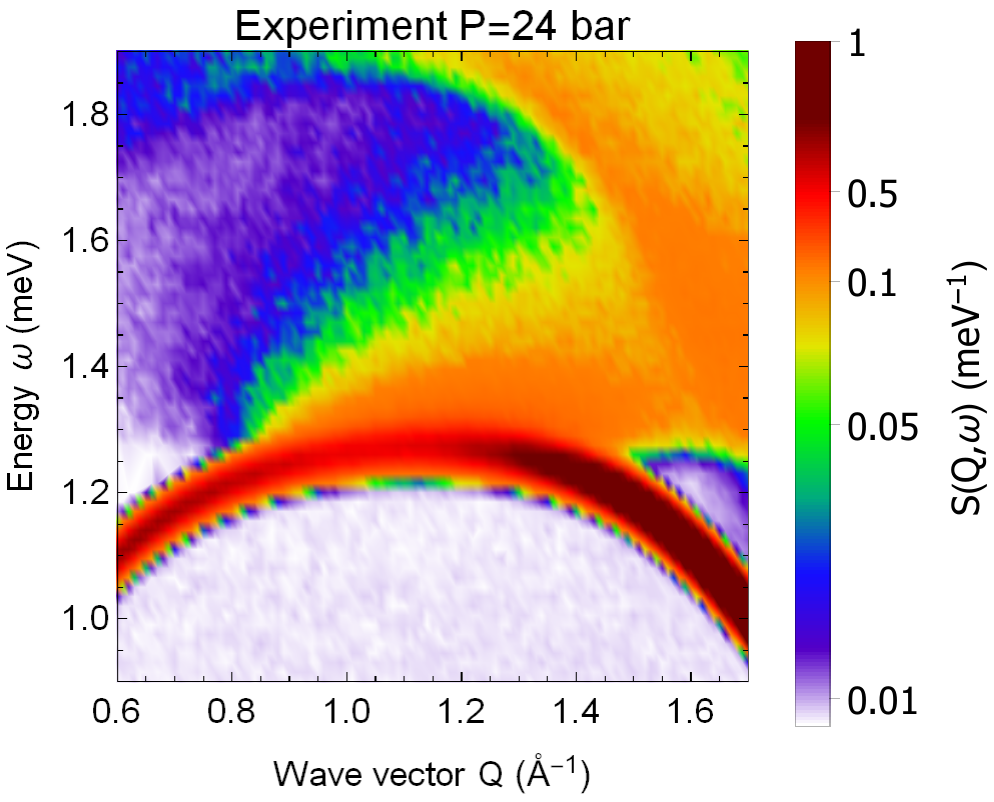}\\
    \vspace{0.3cm}
  \includegraphics[width=0.8\columnwidth]{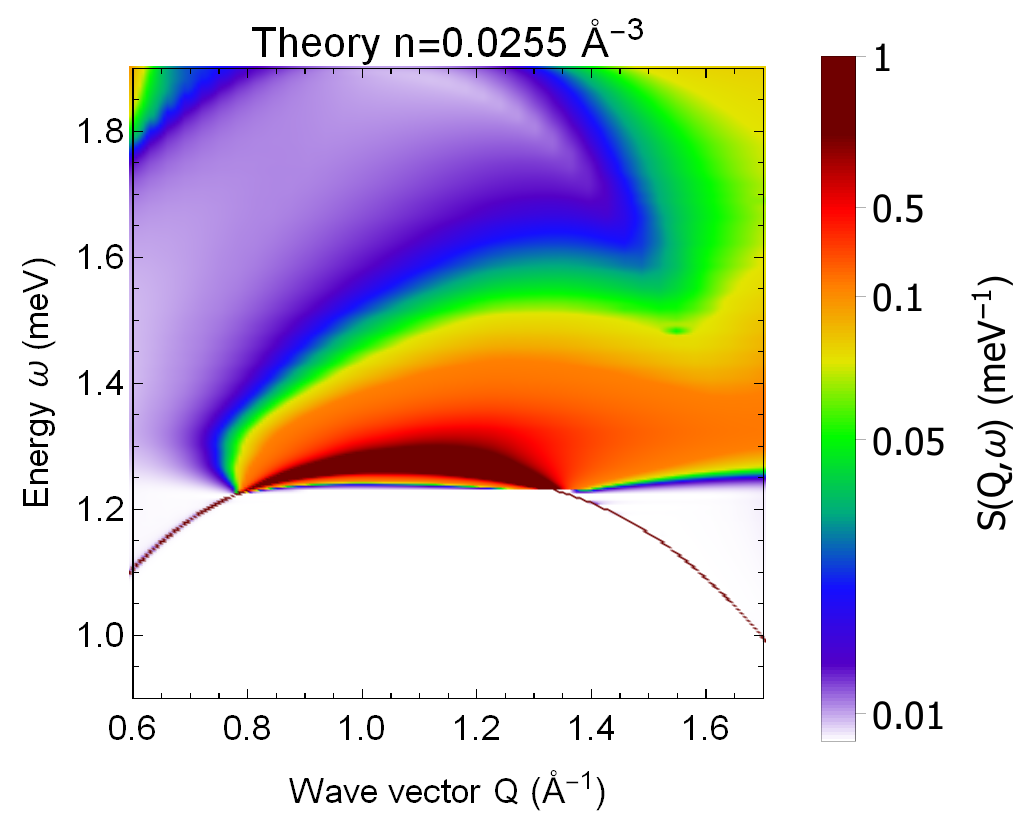}
%\includegraphics[width=0.9\columnwidth]{Fig16a-AboveMaxonP24.pdf}
%\vspace{7cm}
	%\includegraphics[width=0.65\columnwidth]{DensityPlot255aboveMaxon}
  %\includegraphics[width=0.9\columnwidth]{Fig16b-AboveMaxon255.pdf}
  \caption{$S(Q,\omega)$ in the region of the maxon at $P=24$\,bars
    (experiment) and at the corresponding density of 0.0255 \AA$^{-3}$
    (theory).}
		\label{above-maxon} 
\end{figure}

\begin{figure}[h]
	\includegraphics[width=1.0\columnwidth]{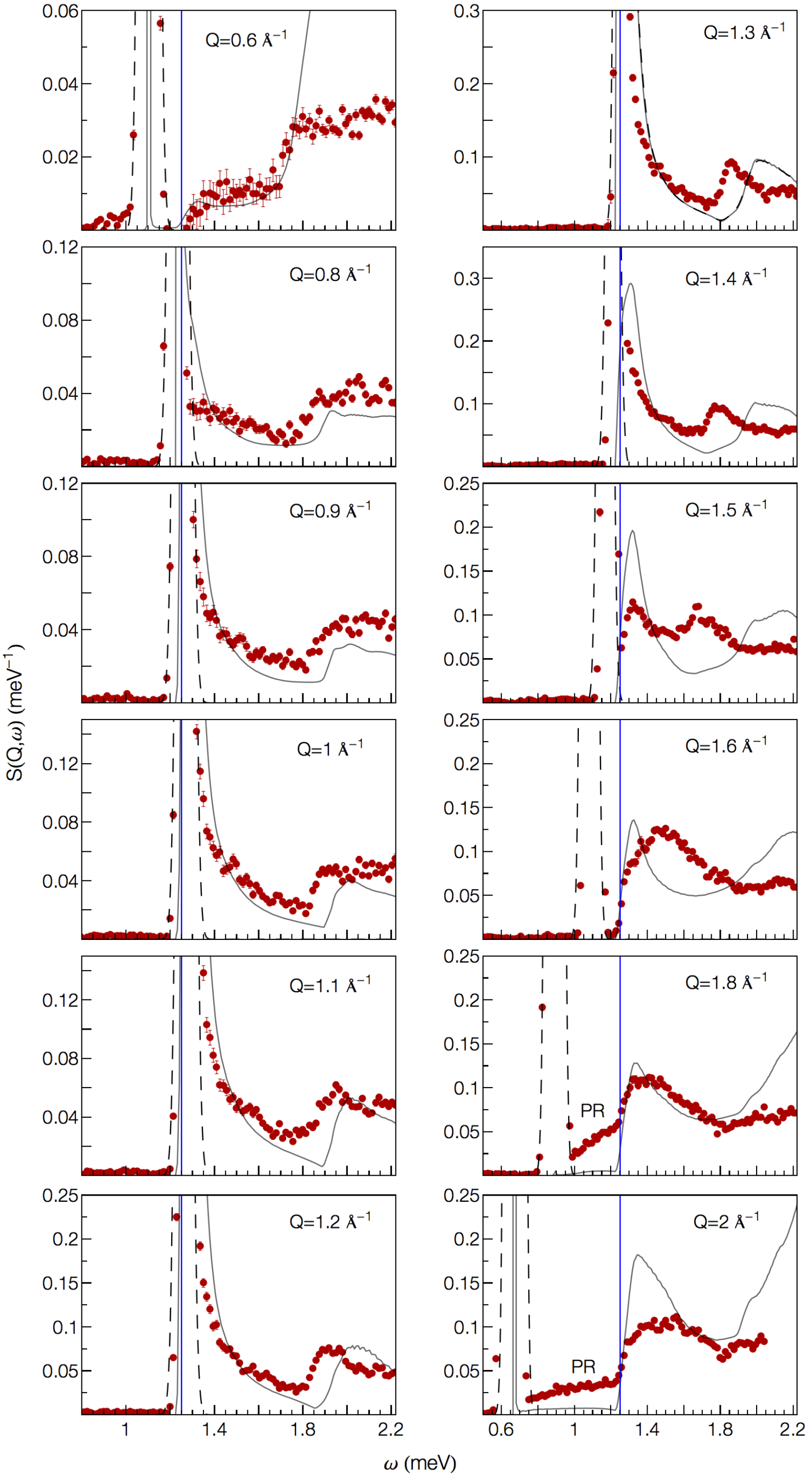}
	\caption{Dynamic structure factor $S(Q,\omega)$: spectra for
          different wave vectors $Q$ in the maxon region, at $P=24$ bars.
          Filled circles: Experimental $S(Q,\omega)$.  Solid lines:
          theory at the density $n$=0.0255 \AA$^{-3}$.  Dashed lines:
          Intensity of the phonon-roton mode (cut off) calculated
          directly from the self energy \cite{eomIII} and convolved
          with the instrumental resolution of 0.07\,meV.  Blue line:
          energy of the roton-roton threshold.  PR indicates
          phonon-roton multi-excitations.}
		\label{CutsP24}
\end{figure}

%\subsubsection{Above the maxon at high pressure}
At high pressures, the maxon energy exceeds twice the roton gap, and a
maxon can decay into two rotons.  We described in Section
\ref{subsec:effect-of-pressure} the broadening of the maxon as it
enters the continuum.  At 24\,bars, the maxon is in the continuum of
the multi-excitations for wave vectors between $Q=0.8$ and $Q=1.5$
\AA$^{-1}$.  Under these conditions, a strong multi-excitation
intensity is observed above the maxon (Figs.\,\ref{above-maxon} and
\ref{CutsP24}).  The very characteristic ``rainbow-like'' measured
spectrum is in very good agreement with the theoretical calculation,
showing in particular that the weight of the maxon is transferred to
the two-roton excitations.

%\subsubsection{The RR-threshold and ``beyond the roton''}
The multi-excitations discussed above, observed above the maxon at
high pressure, are a special case of roton-roton decay.  In fact, a
sharp roton-roton threshold is observed at all wave vectors
(Figs.\,\ref{completemap}, \ref{allpressures} and \ref{CutsP24}), in
regions located far from single-excitations.  The roton-roton
threshold is, in particular, observed at low $Q$ in the present work.
It is also clear, in fact, that the intensity of the RR threshold is
enhanced in the vicinity of single-excitations, as was the case above
the maxon at 24 bars, but also in the region above the Pitaevskii
plateau. Theory and experiment display a similar shape of the spectra
and intensity pattern around the roton-roton threshold, at all
pressures (see Figs.\,\ref{completemap} and \ref{allpressures}).

\subsection{Higher order multi-excitations}

The sharp ``branches'' described above correspond to decay mechanisms
into 2-excitations. Phase-space arguments show that the signal of
higher order processes will be distributed in a rather featureless way
in the energy-wave vector space, due to the vector addition of
momenta. However, the data of Fig.\,\ref{completemap} show that the
multi-excitations region at wave vectors on the order of
1.5\,\AA$^{-1}$ extends to rather high energies, on the order of
4\,meV. This last value constitutes a clear experimental demonstration
that multi-excitations of higher order, related to 3 and 4
single-excitations (the energy of rotons and maxons is on the order of
1\,meV), play a significant role in the dynamics of superfluid $^4$He.

One can also examine the corresponding effect on the wave vector axis,
beyond the end-point of the Pitaevskii plateau.  The plateau could be
expected to end at $2Q{_{R}}$, for a multi-excitation of energy
$2\Delta{_{R}}$ decaying into two rotons of colinear wave
vectors. Previous measurements \cite{fak98,Glyde-EPL-1998,Pearce2001} 
have found that the plateau intensity vanishes at $Q=3.6$ \AA$^{-1}$,
considerably below $2Q{_{R}}=3.84$\,\AA$^{-1}$. This is also observed
in the present work, as seen in Figs.\,\ref{completemap} and
\ref{Cuts}. This effect has been attributed to the decay into two
rotons with an attractive R-R interaction \cite{Pistolesi1998}, 
but other possible interpretations of the data are presently debated. 
We also note that the
intensity does not extend to higher $Q$-values at higher energies as
expected for decays into 3- and 4-excitations processes, an effect
which is probably related to the small phase-space available for
colinear combinations of wave vectors.  As discussed above, the
energy, a scalar, is a better probe for detecting high order
multi-excitation processes. The data at 24\,bars display similar
effects with a simple shift towards higher wave vectors, due to the
larger value of $Q{_{R}}$=2.06\,\AA$^{-1}$ at this pressure.

We now concentrate on the multi-excitations region located slightly
below the free-particle dispersion curve, around 2.5\,\AA$^{-1}$ (see
Fig.\,\ref{completemap}). Earlier studies \cite
{CowleyWoods,Fak91,Pearce2001} observed a rather intense broad feature
extending to higher energies. We find here a much finer structure than
previously believed, and also that it depends rather strongly on the
pressure.  Multi-excitations in this region can only decay into 3 or
more single-excitations, which is therefore of interest for mode-mode
coupling theories. The fact that we observe a high intensity peak is
probably related, at these relatively high energies, to an enhanced
system response in the vicinity of the free-particle dispersion curve,
which is the asymptotic behavior at higher energies. The peak at
24\,bars is less intense than the corresponding one at SVP, suggesting
that the maxon, strongly reduced at this pressure, is involved in the
corresponding decay processes.

Finally, at the highest energies explored here, $S(Q,\omega)$
progressively converges towards the free-particle parabola, remaining
below it (see Fig.\,\ref{completemap}). The so-called ``glory''
oscillations seen as a function of $Q$, both in the peak position and
the width, are well documented in the literature
\cite{AndersenPRB1997}. Directly related to the corresponding
oscillations in the static structure factor $S(Q)$, they result from the
hard core part of the $^4$He-$^4$He interaction potential and from
quantum coherence effects. Earlier works could not fit the spectra of
the first oscillation with a single peak. The highly structured
multi-excitations seen in the present work show that this peak of
unusual shape results in fact from the superposition of a few
multi-excitation ``branches'' corresponding to decays into a few
single-excitations. Again, the dynamic structure factor in this region
depends on pressure, and the spectra for $Q\approx{3.5}$\,\AA$^{-1}$
are strongly affected by the collapse of the maxon.

\section{Conclusion}

A comprehensive understanding of the dynamics of interacting Bose
systems, going from the Landau quasi-particles and multi-excitations
regimes, up to the high-energy limit where the independent particle
dynamics is recovered, emerges from our combined experimental 
and theoretical work. 
The up-to-now largely unexplored multi-excitations regime has been
systematically investigated. Ghost-phonon and ghost-roton regimes have
been observed, associated to phonon emission in the region of nearly
supersonic multi-excitations, by a Cherenkov-like process
qualitatively predicted by Burkova's extension of Pitaevskii's
theory. Several other multi-excitation branches or thresholds have
been observed and identified in the low energy sector, where an
excellent quantitative agreement is found with the predictions of the
Dynamic Many-Body theory. This agreement extends even to high
pressures, near solidification, as shown for example for the
remarkable case of the maxon disintegration into two rotons.  
The calculations including specific multiparticle 
fluctuations to all orders \cite{eomIII} provide
a good description of the dynamics for energies as high as 2\,meV.
Above this value, higher order processes dominate the dynamics. Our
high energy/wave vector data call for further theoretical developments
able to describe quantitatively the behavior observed at higher
energies, above the simple multi-excitations region but still
substantially below the quasi-free particle (impulse-approximation)
sector.

\section{Acknowledgements}
We are grateful to X. Tonon for his help with the experiments.
This work was supported, in part, by the Austrian Science Fund FWF grant I602, 
the French grant ANR-2010-INTB-403-01, the European Community Research 
Infrastructures under the FP7 Capacities Specific Programme, Microkelvin  
project number 228464, and the European Microkelvin Platform.
 
%\newpage

%\bibliography {Pressure-v9c}
\bibliographystyle{apsrev4-1}

\end{document}